\RequirePackage{fix-cm}

\documentclass[peerreview]{IEEEtran}
\usepackage{amsthm}
\usepackage[margin=1.2in]{geometry}

\usepackage{graphicx}
\usepackage{amssymb}
\usepackage{amsmath}

\newtheorem{prop}{Proposition}

\newtheorem{lem}{Lemma}

\begin{document}

\title{Track estimation with binary derivative observations
}


\author{Adrien Ickowicz \thanks{A. Ickowicz, CEREMADE, University of Paris-Dauphine, Tel.: +33-1-45465874, e-mail: ickowicz@ceremade.dauphine.fr}         
           }

\date{Received: date / Accepted: date}

\maketitle

\begin{abstract}
We focus in this paper in the estimation of a target trajectory defined by whether a time constant parameter in a simple stochastic process or a random walk with binary observations. The binary observation comes from binary derivative sensors, that is, the target is getting closer or moving away. Such a binary obervation has a time property that will be used to ensure the quality of a max-likelihood estimation, through single index model or classification for the constant velocity movement. In the second part of this paper we present a new algorithm for target tracking within a binary sensor network when the target trajectory is assumed to be modeled by a random walk. For a given target, this algorithm provides an estimation of its velocity and its position. The greatest improvements are made through a position correction and velocity analysis. 
\end{abstract}

\section{Introduction}

Sensor networks are systems made of many small and simple sensors deployed over an area in an attempt to sense events of interest within that particular area. In general, the sensors have limited capacities in terms of say range, precision, etc. The ultimate information level for a sensor is a binary one, referring to its output. However, it is important to make a distinction according to the nature of this binary information. Actually, it can be related to a $0-1$ information (non-detection or detection) or to relative $\{-,+\}$ motion information. For example, if the sensors are getting sound levels, instead of using the real sound level (which may cause confusion between loud near objects and quieter close objects), the sensor may simply report whether the Doppler frequency is suddenly changing, which can be easily translated in whether the target is getting closer or moving away. Moreover, low-power sensors with limited computation and communication capabilities can only perform binary detection. We could also cite video sensors, with the intuitive reasoning: the target is getting closer if its size is increasing. The need to use that kind of sensor networks leads to the development of a model for target tracking in binary sensor networks.\\

\begin{figure}[ht!]
{\center
{\includegraphics[width=6cm, height=4cm]{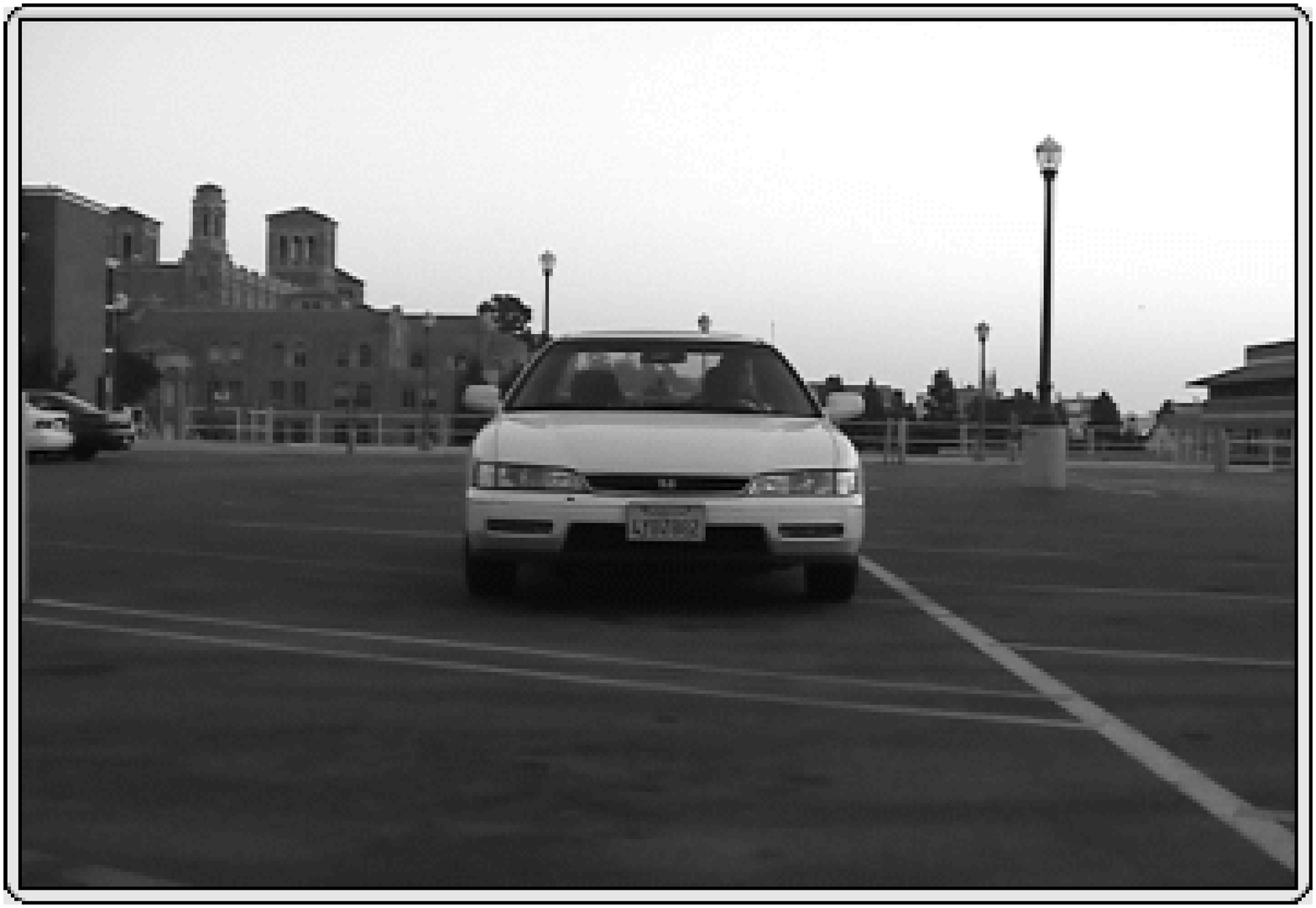}}
{\includegraphics[width=6cm, height=4cm]{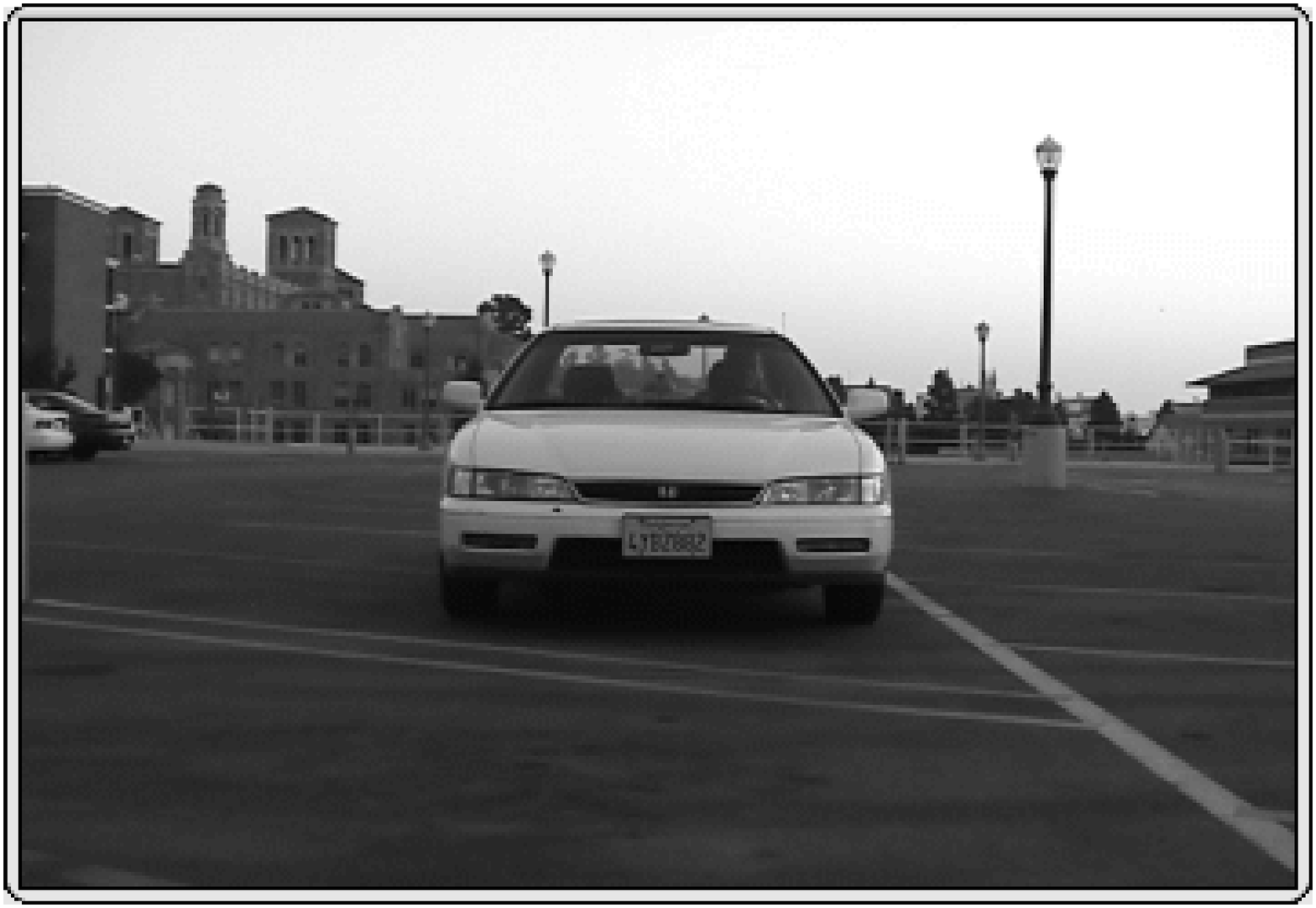}}
\caption{\it A view of a getting closer car}}
\label{sensorsh}
\end{figure}

We consider a sensor network, made with $N$ sensors (e.g. video),with (known) positions. Each sensor can only gives us a binary $\{-,+\}$ information \cite{asl}, i.e. whether the target-sensor distance is decreasing ($-$) or increasing ($+$).  This "choice" can result from severe communication requirements or from the difficulties from fusing inhomogeneous data.  Even if many important  works deal with proximity sensors \cite{Wan}, \cite{laz}, we decide here to focus on the  binary $\{-,+\}$ information \cite{asl}. Here, the aim is  to estimate the parameters defining the target trajectory. Even if our methods can be rather easily extended  to more complex models of target motion, we decide to focus here  on a constant velocity movement. Actually, this framework is sufficiently general to present the main problems we have to face, as well as the foundations of the methods we have to develop for dealing with these binary data. See fig. \ref{sensorsh} for an example.\\
In a first time, the observability requirements are considered. Then, we turn toward the development of specific estimation methods. Especially, the new concept of the velocity plane is introduced as an exhaustive representation of the spatio-temporal sequence of binary data. It is then used both in a separation-oriented framework (SVM) and in a projection pursuit regression (PPR) one. The corresponding methods are carefully presented and analyzed. 

In the following part we release the assumption of (piecewise) constant velocity motion, and we try to follow both position and velocity in real time. In particular, it is shown that it is the trajectory "diversity" which renders this possible. \\

Obviously, tracking a diffusive Markovian target widely differs from the (batch) estimation of deterministic parameters. However, both problems present strong similarities. Indeed, the geometrical properties remains the same at each instant. Once the target motion model has been introduced, the most important properties we used to perform the tracking are presented. Then, the method which allows us to perform adapted corrections for tracking the target is presented. It is the main contribution of this part of the paper. 

Simulation results illustrate the behavior of the estimators, as well as the performances of the tracking algorithm. We conclude on further works about the tracking in binary sensor networks.

\section{Binary Sensor Network Observability Properties}
\label{obs}

Let us denote ${\sf s}_{i}$ a sensor whose position is represented by the the vector ${\bf t}_{i}$ Similarly, the vector ${\bf x}_{t}$ represents the position vector of the target at the time-period $t$.  Let us denote $d_{i}(t)$ the (time-varying) distance from sensor ${\sf s}_{i}$ to the target  at time $t$. Then, we have that:
\begin{equation}
d_{i}(t)\;\searrow \Longleftrightarrow  {\dot{d}}_{i}(t)<0\;,\mbox{or:~}\;\langle {\bf x}_{t}-{\bf t}_{i}, {\bf v}_{t} \rangle <0\;,
\end{equation}
where ${\bf v}_{t}$ is the instantaneous target velocity. We thus have the following lemma.
\begin{lem}
Let ${\sf s}_{i}$ (resp. ${\sf s}_{j}$) a sensor whose the target distance is decreasing  (resp. increasing) at the time-period $t$, then we have:
\begin{equation}
\langle {\bf t}_{j},{\bf v}_{t}\rangle \;< \;\langle {\bf x}_{_t},{\bf v}_{t} \rangle\;<\;\langle {\bf t}_{i},{\bf v}_{t}\rangle  \;.
\label{+-1}
\end{equation}
\label{+-lem}
\end{lem}

If we restrict to binary motion information, we consider that the output $s_{i}(t)$  of a sensor (at time $t$) is $+1$ or $-1$ according to the distance $d_{i}(t)$ is decreasing or increasing, so that we have:
\begin{equation}
\left\{
\begin{array}{l}
\medskip
s_{i}(t)=+1\;\;\mbox{if}\;\;{\dot{d}}_{i}(t)<0\;,\\
s_{j}(t)=-1\;\;\mbox{if}\;\;{\dot{d}}_{j}(t)>0\;.
\end{array}
\right.
\end{equation}
Let us denote $A$ the subset of sensor whose output is $+1$ and $B$ the subset of sensors whose output is $-1$, i.e. $A=\left\{ {\sf s}_{i}|\;s_{i}(t)=+1 \right\}$ and $B=\left\{ {\sf s}_{j}|\;s_{j}(t)=-1 \right\}$ and $C(A)$ and $C(B)$ their convex hulls, then we have \cite{asl}:
\begin{prop}
$C(A) \bigcap C(B)=\emptyset$ and ${\bf x}_{t} \notin C(A)\bigcup C(B)$.
\end{prop}

{\bf Proof:} The proof is quite simple is reproduced here only for the sake of completeness. First assume that $C(A)\bigcap C(B)\neq \emptyset$, this means that there exists an element of $C(B)$, lying in $C(A)$. Let ${\sf s}$ be this element (and ${\bf t}$ its associated position), then we have ($t \in C(B)$):
\begin{equation}
\begin{array}{l}
{\bf t}=\displaystyle{\sum_{j\in B}} \beta_{j}\;{\bf t}_{j}\;,\;\beta_{j}\geq 0\;\mbox{and~}\;\displaystyle{\sum_{j\in B} }\beta_{j}=1\,\\
\mbox{so that we have on the first hand:}\\
\langle {\bf t},{\bf v}_{t} \rangle= \displaystyle{\sum_{j\in B}} \beta_{j}\;\langle {\bf t}_{j}, {\bf v}_{t}\rangle\;<\;\langle {\bf x}_{t},{\bf v}_{t} \rangle \;\;(\mbox{see eq. \ref{+-1}}),\\
\mbox{and, on the other one ($t \in C(A)$):}\\
\langle {\bf t},{\bf v}_{t} \rangle=\displaystyle{\sum_{i\in A}} \alpha_{i}\;\langle {\bf t}_{i}, {\bf v}_{t}\rangle\;\geq \left(\displaystyle{\sum_{i\in A} } \alpha_{i}\right)\;\displaystyle{\min_{i}}\;\left\{\langle {\bf t}_{i}, {\bf v}(t) \rangle \right\}>\langle {\bf x}_{t},{\bf v}_{t} \rangle\;.
\end{array}
\end{equation}
Thus a contradiction which shows that $C(A) \bigcap C(B)=\emptyset$. For the second part, we have simply to assume that ${\bf x}(t) \in C(A)$ (~${\bf x}_{t}= \displaystyle{\sum_{i\in A} }\alpha_{i} \;{\bf t}_{i},\;\alpha_{i}\geq 0$), which yields:
\begin{equation}
\langle {\bf x}_{t},{\bf v}_{t} \rangle =\displaystyle{\sum_{i\in A}} \alpha_{i} \;\langle {\bf t}_{i},{\bf v}_{t}\rangle \geq \displaystyle{\min_{i \in A}} \langle {\bf t}_{i},{\bf v}_{t} \rangle,
\end{equation}
which is clearly a contradiction, idem if $X(t) \in C(B)$. \\
$\Box\Box\Box$\\

So, $C(A)$ and $C(B)$ being two disjoint convex subsets, we know that there exists an hyperplane (here a line) separating them. 
Then, let ${\sf s}_{k}$ be a generic sensor, we can write ${\bf t}_{k}=\lambda\;{\bf v}_{t}+\mu\;{\bf v}_{t}^{\perp}$, so that:
\begin{equation}
\langle {\bf t}_{k},{\bf v}_{t} \rangle=\lambda \,{\|{\bf v}_{t}\|}^{2}>0  \Longleftrightarrow \lambda>0\;.
\end{equation}
This means that the line spanned by the vector ${\bf v}_{t}^{\perp}$ separates $C(A)$ and $C(B)$. Without considering the translation and considering again the $\left\{ {\bf v}_{t},{\bf v}_{t}^{\perp} \right\}$ basis , we have :
\begin{equation}
\left\{
\begin{array}{l}
{\bf t}_{k} \in A \Longleftrightarrow  \lambda\; {\| {\bf v}_{t} \|}^{2}>\langle {\bf x}_{t},{\bf v}_{t} \rangle \;,\\
{\bf t}_{k} \in B \Longleftrightarrow  \lambda \;{\| {\bf v}_{t} \|}^{2}<\langle {\bf x}_{t},{\bf v}_{t} \rangle \;.
\end{array}
\right.
\end{equation}
Thus in the basis $({\bf v}_{t}, {\bf v}_{t}^{\perp})$, the line passing by the point $\displaystyle{\left(\frac{ \langle {\bf x}_{t},{\bf v}_{t} \rangle} {{\| {\bf v}_{t} \|}^{2} },0\right)}$ and whose direction is given by ${\bf v}_{t}^{\perp}$ is separating $C(A)$ and $C(B)$. 
 We have now to turn toward the indistinguishability conditions for two trajectories. Two trajectories are said indistinguishable if they induce the same outputs from the sensor network. We have then the following property \cite{asl}.

\begin{prop}
Assume that the sensor network is {\it dense}, then two target trajectories (say ${\bf x}_{t}$ and ${\bf y}_{t}$)  are indistinguishable iff the following conditions hold true:
\begin{equation}
\left\{
\begin{array}{l}
\medskip
{\dot{\bf y}}_{t}=\lambda_{t}\;{\dot{\bf x}}_{t} \;\; (\lambda_{t}>0) \;\;\forall t\;,\\
\langle {\bf x}_{t}-{\bf y}_{t}, {\dot{\bf x}}_{t} \rangle =0 \;\;\forall t\;.
\end{array}
\right.
\end{equation}
\label{inftysens}
\end{prop}

{\bf Proof:~}~First, we shall consider the implications of the indistinguishability. Actually, the two trajectories are indistinguishable iff the following condition holds:
\begin{equation}
\langle {\bf t}_{j}-{\bf t}_{i}, {\dot{\bf x}}_{t} \rangle \leq 0 \Longleftrightarrow \langle {\bf t}_{j}-{\bf t}_{i}, {\dot{\bf y}}_{t}\rangle  \leq 0\;\;\forall t\;\;\forall ({\bf t}_{i},{\bf t}_{j})\;.
\label{titj}
\end{equation}
We then {\it choose} ${\bf t}_{j}-{\bf t}_{i}=\alpha \;{\dot{\bf x}}_{t}^{\perp}$ (i.e. ${\bf t}_{i}$ and  ${\bf t}_{j}$ both belongs to the line separating $A$ and $B$) and consider the following decomposition of the ${\dot{\bf y}}_{t}$ vector:
$$
{\dot{\bf y}}_{t}=\lambda_{t}\; {\dot{\bf x}}_{t}+\mu_{t}\;{\dot{\bf x}}_{t}^{\perp}\;,
$$
so that we have:
\begin{equation}
\langle {\bf t}_{j}-{\bf t}_{i}, {\dot{\bf y}}_{t} \rangle =\alpha\mu_{t} \;{\| {\dot{\bf x}}_{t}^{\perp} \|}^{2} \leq 0\;.
\end{equation}
Now, it is always possible to choose a scalar $\alpha$ of the same sign than $\mu_{t}$. So, we conclude that the scalar $\mu_{t}$ is necessarily equal to zero. Thus , if the trajectories ${\bf x}_{t}$ and ${\bf y}_{t}$ are indistinguishable we have necessarily:
$$
{\dot{\bf y}}_{t}=\lambda_{t}\;{\dot{\bf x}}_{t} \;,\;\forall t\;.
$$
Furthermore, the scalar $\lambda_{t}$ is necessarily positive (see eq. \ref{titj}). Then, the lemma \ref{+-lem} inequalities yield:
\begin{equation}
\langle {\bf t}_{j}-{\bf t}_{i}, {\dot{\bf x}}_{t} \rangle \;<\;\langle {\bf x}_{t}-{\bf y}_{t}, {\dot{\bf x}}_{t} \rangle\;<\;\langle {\bf t}_{i}-{\bf t}_{j}, {\dot{\bf x}}_{t} \rangle \;.
\label{ineq}
\end{equation}
Choosing once again ${\bf t}_{j}-{\bf t}_{i}=\alpha \;{\dot{\bf x}}_{t}^{\perp}$, we deduce from eq. \ref{ineq} the second part of prop. \ref{inftysens}, i.e. $\langle {\bf x}_{t}-{\bf y}_{t}, {\dot{\bf x}}_{t} \rangle =0 \;\;\forall t\;$. Considering now the distance between the two indistinguishable trajectories, we have (${\dot{\bf y}}_{t}=\lambda_{t}\; {\dot{\bf x}}_{t}$) :
\begin{equation}
\frac{d}{dt}\; {\|{\bf x}_{t}-{\bf y}_{t}\|}^{2}=2\;\langle {\bf x}_{t}-{\bf y}_{t}, {\dot{\bf x}}_{t}- {\dot{\bf y}}_{t}\rangle=0\;,
\end{equation}
so that we have $\|{\bf x}_{t}-{\bf y}_{t}\|={\sf cst}$.\\
 Reciprocally, assume that the two conditions ${\dot{\bf y}}_{t}=\lambda_{t}\;{\dot{\bf x}}_{t}$ and $\langle {\bf x}_{t}-{\bf y}_{t}, {\dot{\bf x}}_{t} \rangle =0 $ hold true $\forall t$, are the two trajectories then indistinguishable? It is sufficient to remark that:
\begin{equation}
\begin{array}{l}
\medskip
\langle {\bf y}_{t}, {\dot{\bf y}}_{t} \rangle =\langle {\bf x}_{t}+({\bf y}_{t}-{\bf x}_{t}),{\dot{\bf y}}_{t} \rangle=\langle {\bf x}_{t}, {\dot{\bf y}}_{t}\rangle=\lambda_{t}\; \langle {\bf x}_{t}, {\dot{\bf x}}_{t} \rangle \;,\\
\langle {\bf t}_{i}, {\dot{\bf y}}_{t} \rangle =\lambda_{t}\; \langle {\bf t}_{i}, {\dot{\bf x}}_{t} \rangle \;.
\end{array}
\end{equation}
Since the scalar $\lambda_{t}$ is positive this ends the proof. \\
$\Box\Box\Box$\\

Let us now consider the practical applications of the above general results. 

\subsection*{\bf Rectilinear and uniform motion}

Admitting now that the target motions are rectilinear and uniform (i.e. ${\bf x}_{t}={\bf x}_{0}+t \;\dot{\bf x}$). Then prop. \ref{inftysens} yields $\dot{\bf y}=\lambda \;\dot{\bf x}$~ ($\lambda>0$) and:
\begin{equation}
\langle {\bf y}_{t}-{\bf x}_{t}, \dot{\bf x} \rangle=\langle {\bf y}_{0}-{\bf x}_{0}, \dot{\bf x} \rangle+t \;(1-\lambda)\;{\|\dot{\bf x} \|}^{2}=0\;\forall t\;.
\label{toute}
\end{equation} 
Then, from eq. \ref{toute} we deduce that $\lambda=1$ and ${\bf y}_{0}={\bf x}_{0}+\alpha\;{\dot{\bf x} }^{\perp}$. So that, the target velocity is fully observable while the position is uniquely determined modulo a $\alpha\;{\dot{\bf x} }^{\perp}$ translation.

\subsection*{leg-by-leg trajectory}

 Consider now a leg-by-leg trajectory modeling. For a $2$-leg one, we have for two indistinguishable trajectories:
\begin{equation}
\left\{
\begin{array}{l}
{\bf x}_{t}={\bf x}_{0}+t_{1}\;{\bf v}_{x}^{1}+(t-t_{1})\;{\bf v}_{x}^{2}\;,\\
{\bf y}_{t}={\bf y}_{0}+t_{1}^{'}\;{\bf v}_{y}^{1}+(t-t_{1}^{'})\;{\bf v}_{y}^{2}\;,
\end{array}
\right.
\label{leg-leg1}
\end{equation}
where ${\bf v}_{x}^{i}$ is the velocity of the ${\bf x}(t)$ trajectory on the $i$-th leg and $t_{i}$ is the epoch of maneuver. Furthermore, we can assume that $t_{1}<t_{1}^{'}$. Considering the implications of prop. \ref{inftysens} both for $t<t_{1}$ and for $t>t_{1}^{'}$, we know that if the trajectories are indistinguishable we must have:
\begin{equation}
{\bf v}_{x}^{1}={\bf v}_{y}^{1} \;\;\mbox{and:~}\;{\bf v}_{x}^{2}={\bf v}_{y}^{2}\;.
\end{equation}
So, our objective is now to prove that we have also $t_{1}=t_{1}^{'}$. 
Considering prop. \ref{inftysens}, we thus have the following system of equations :
\begin{equation}
\left\{
\begin{array}{llll}
 \langle {\bf y}_{0}-{\bf x}_{0} +(t-t_{1})\;\left({\bf v}_{x}^{1}-{\bf v}_{x}^{2}\right), {\bf v}_{x}^{1} \rangle=0 &\;\mbox{for~:} & t_{1}<t<t_{1}^{'}&\;(a),\\
\langle {\bf y}_{0}-{\bf x}_{0} +(t-t_{1})\;\left({\bf v}_{x}^{1}-{\bf v}_{x}^{2}\right), {\bf v}_{x}^{2} \rangle=0 & \;\mbox{for~:} & t_{1}<t<t_{1}^{'}& \;(b),\\
\langle {\bf y}_{0}-{\bf x}_{0} +(t_{1}^{'}-t_{1})\;\left({\bf v}_{x}^{1}-{\bf v}_{x}^{2}\right),{\bf v}_{x}^{2}\rangle =0 & \;\mbox{for~:}& t_{1}^{'}<t& \;(c)\;.
\end{array}
\label{leg-leg3}
\right.
\end{equation}
Now, on the $1$-st leg we have also $\langle {\bf y}_{0}-{\bf x}_{0}, {\bf v}_{x}^{1} \rangle=0$ (see prop. \ref{inftysens} for $t=0$), so that eqs~ \ref{leg-leg3}a,b yield:
\begin{equation}
\langle \left({\bf v}_{x}^{1}-{\bf v}_{x}^{2}\right), {\bf v}_{x}^{1} \rangle= \langle \left({\bf v}_{x}^{1}-{\bf v}_{x}^{2}\right), {\bf v}_{x}^{1} \rangle=0\;.
\label{leg-leg4}
\end{equation}
This means that ${\bf v}_{x}^{1}$ and ${\bf v}_{x}^{2}$ are both orthogonal to the same vector (${\bf v}_{x}^{1}-{\bf v}_{x}^{2}$), so they are collinear, and we straightforwardly deduce from eq. \ref{leg-leg4} that ${\bf v}_{x}^{1}={\bf v}_{x}^{2}$. Finally, it has thus been proved that $t_{1}=t_{1}^{'}$ and this reasoning can be extended to any leg number. The observability requirements having been considered, we turn now toward the development of the algorithmic approaches. Let us first introduce the following functional.

\section{The stairwise functional}
\label{stairwise}

\begin{figure}[ht!]
{\includegraphics[width=9cm, height=8cm]{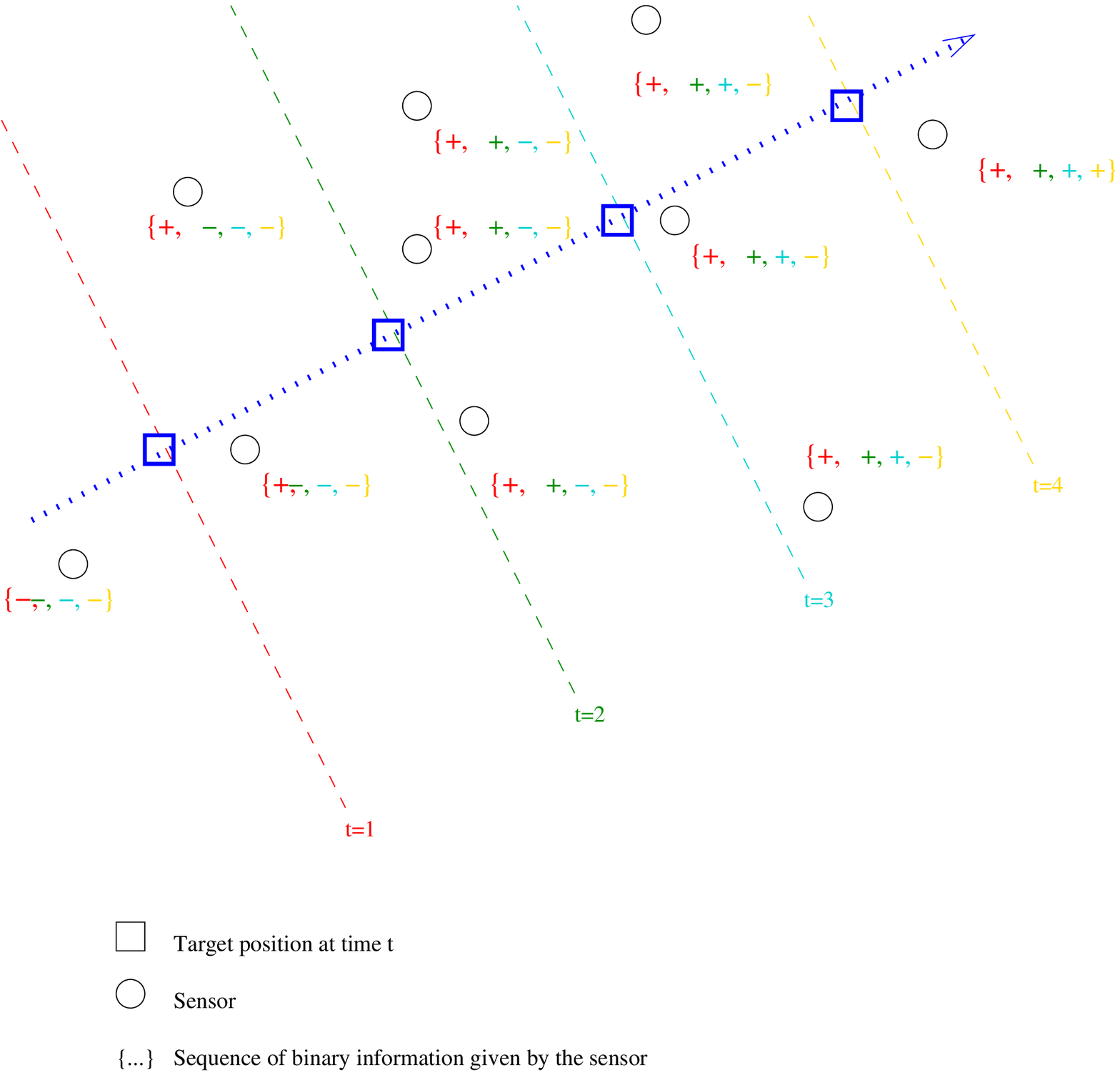}}
{\includegraphics[width=9cm, height=8cm]{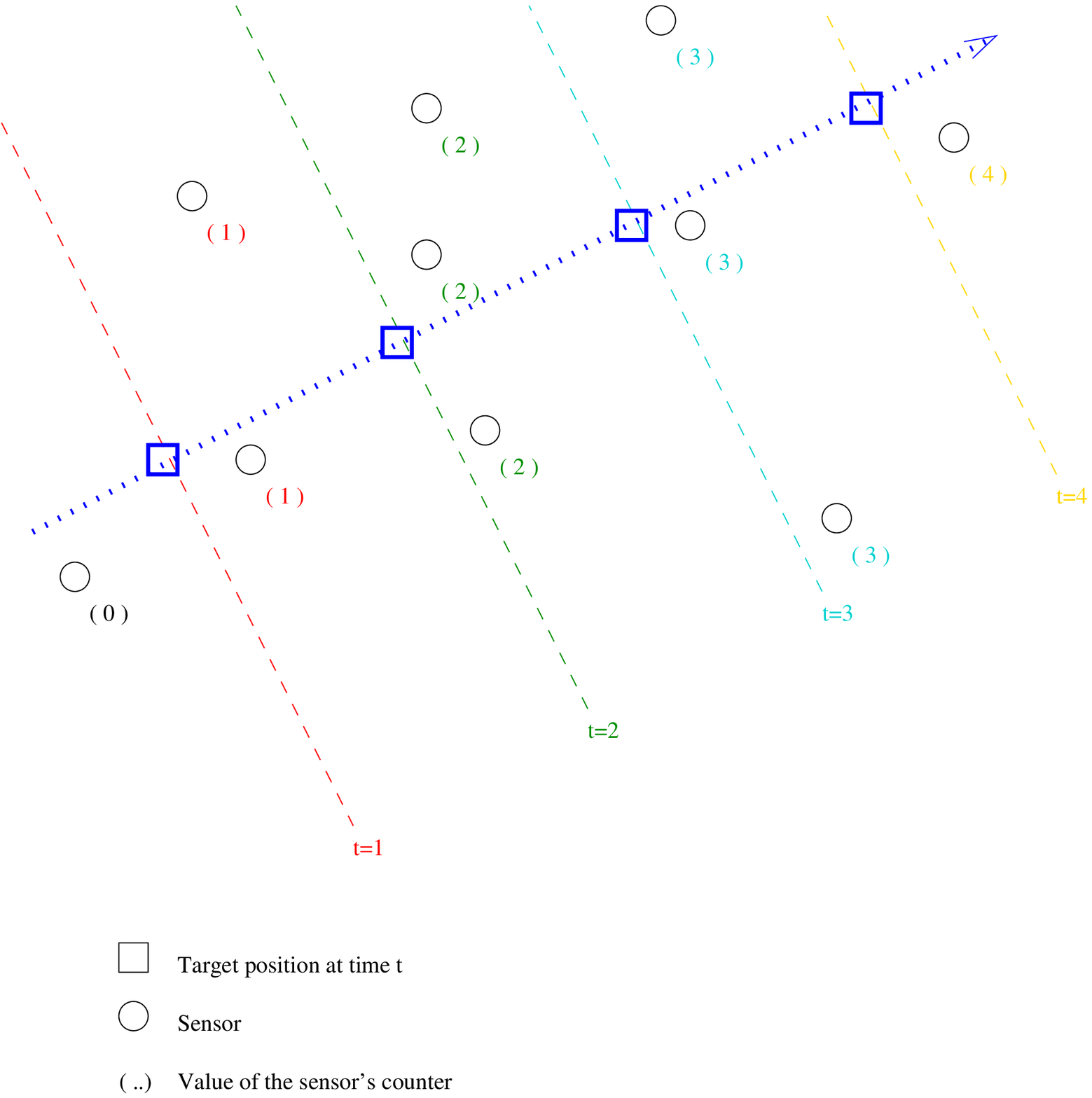}}
\caption{\it A scenario of target evolution and sensor network information}
\label{network1}
\end{figure}

\begin{figure}[ht!]
{\includegraphics[width=9cm, height=8cm]{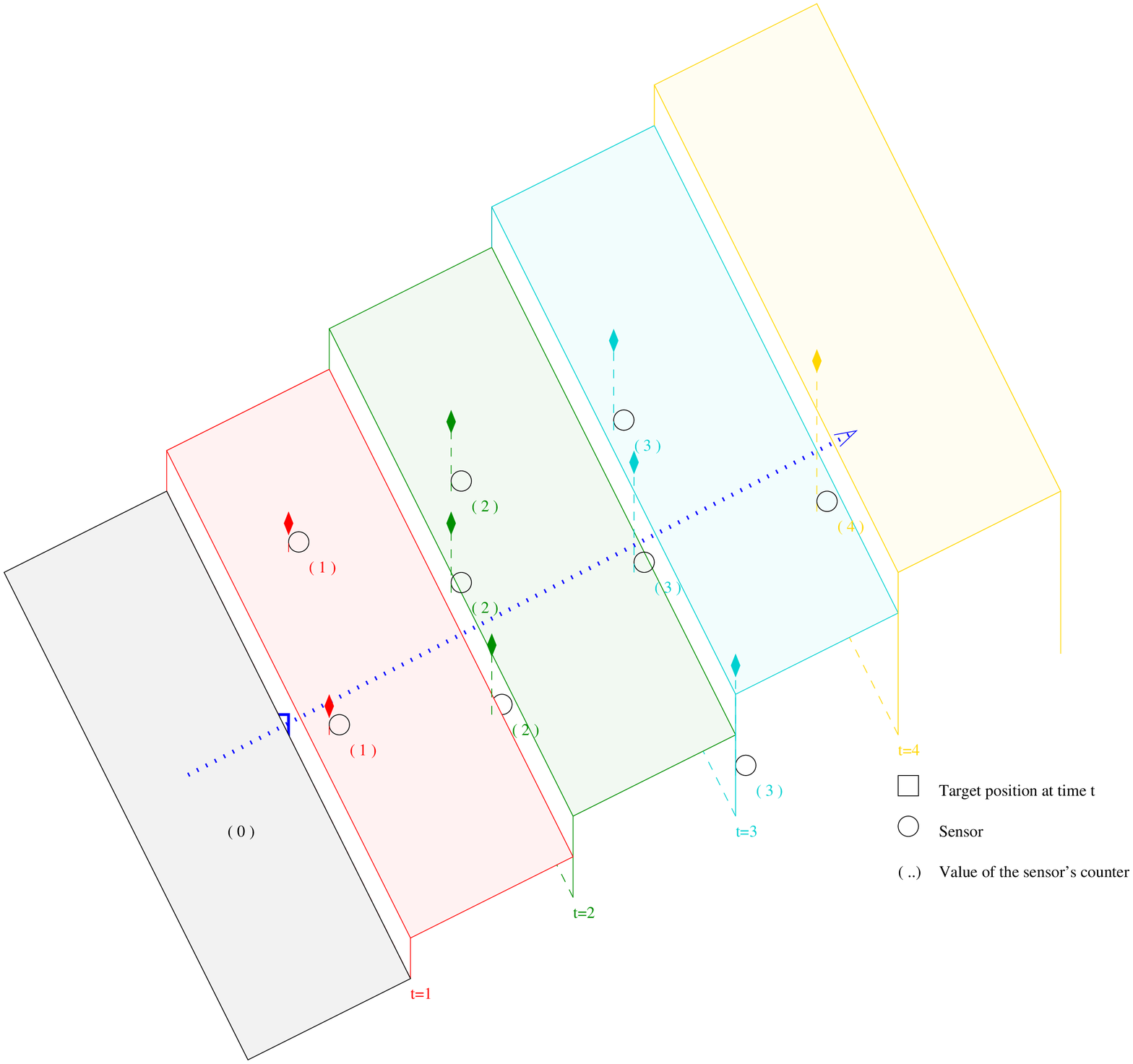}}
{\includegraphics[width=9cm, height=8cm]{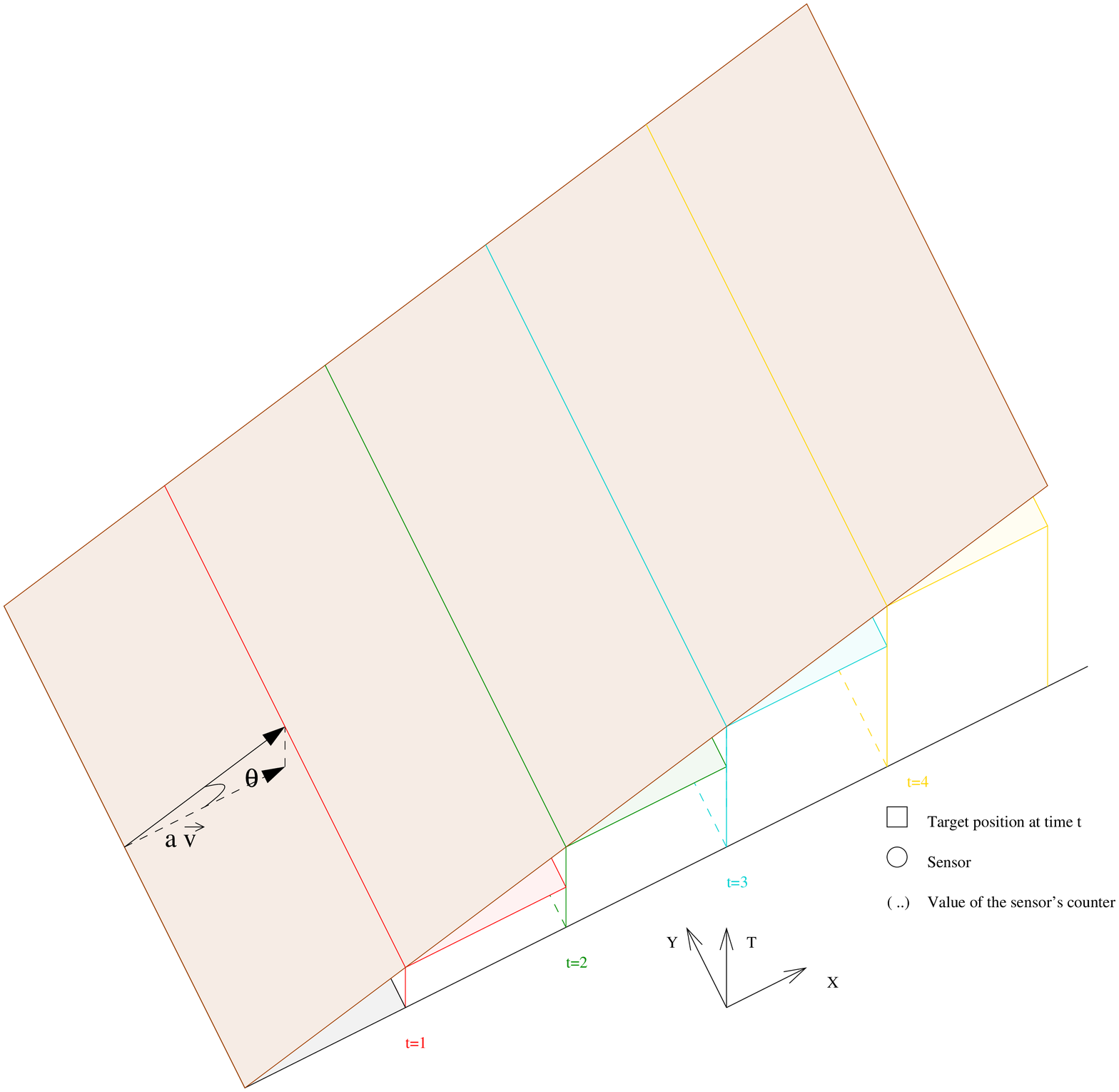}}
\caption{\it The theoretical stairway of the trajectory.}
\label{stair1}
\end{figure}

Our fist aim is to estimate the target velocity, within a batch processing framework. We assume that $N$ binary ($\left\{-,+ \right\}$ sensors are uniformly distributed on the field of interest (see fig. \ref{network1}).\\
Each sensor will be coupled with a counter, that will be increased by a {\it unity} each time-period the sensor gives us a $\{+\}$, an will keep its value each time the sensor gives us a $\{-\}$. Then, at the end of the trajectory, each sensor has a entire value representing the number of periods the target was approaching. Within a given batch, the outputs of the sensor counters can be represented by a stairwise functional (see fig. \ref{stair1}).\\
Then, once this stair is built, we can define what we call the velocity plane. This plane is the tangent plane of the stairwise functional, which means that its direction gives the direction of the stair, while its angle $\theta$ gives the slope.  The direction of the plane gives us the target heading, while the target speed $v$ is given by:
\begin{equation}
v= \frac{1}{\tan(\theta)}\;.
\end{equation}
Thus, estimating the velocity is equivalent to estimating the velocity plane parameters. Mathematical justifications are then presented.
The target moves with a constant velocity $\bf{v}$. Considering the results of section \ref{obs}, its starting position is given by the following equation:
\begin{equation}
\begin{array}{l}
{\bf x}(0) = {\bf x}_{0} + \lambda \;\bf{v}^{\perp}\;\;,\; \lambda \in \mathbb{R}\;,\\
\mbox{so, that}:\\
{\bf x}(t) ={\bf x}_{0} + \lambda \;\bf{v}^{\perp} + t \;{\bf v}
\end{array}
\end{equation}
This means that at each time period $t \in \mathbb{R}_{+}$, the possible positions ${\bf{x}}(t)$ define a (moving) straight line, whose direction is $\bf{v}^{\perp}$. Let us consider now the scalar product $\langle {\bf x}(t), {\bf v} \rangle $, then we have:
\begin{equation}
\frac{\partial}{\partial \;t} \;\langle {\bf x}(t), {\bf v} \rangle=v^{2}\;.
\end{equation}
This is clearly constant, which means that the surface is a plane. The conclusion follows: the stairwise plane is an exhaustive information for the velocity vector. We provide in the next section two solutions to estimate the velocity plane from the observed data, and  give some asymptotic results about the estimation.

\section{Statistical Methods to Estimate the Velocity Plane}

We showed that estimating the velocity plane allows us to estimate the velocity vector.  Wile there exists several methods to do that, we shall focus on two of them. 

\subsection{The Support Vector Machine (SVM) approach \cite{cor:vap}}

As seen previously, the problem we have to face is to optimally separate the two classes of sensors (i.e. the $+$ and $-$). So, we can use the general framework of SVM, widely used in the  classification context. The set of labeled patterns $\left\{(y_{1},{\bf x}_{1}, \cdots, y_{l},{\bf x}_{l} \right\}$ ($y_{i}\in \left\{-1,1\right\}$ and ${\bf x}_{i}$ sensor positions) is said to be linearly separable if there exists a vector ${\bf w}$ and a scalar $b$ such that the following inequalities hold true:
\begin{equation}
\left\{
\begin{array}{l}
\medskip
\langle {\bf w},{\bf x}_{i} \rangle +b \geq 1 \;\;\;\mbox{if~:~} y_{i}=1\;,\\
\langle {\bf w},{\bf x}_{i} \rangle +b \leq -1 \;\;\;\mbox{if~:~} y_{i}=-1\;.
\end{array}
\right.
\label{svm1}
\end{equation}
Let ${\cal{H}}({\bf w},b)\stackrel{\Delta}{=}\left\{{\bf x}|\langle {\bf w}, {\bf x} \rangle +b=0 \right\}$ (${\bf w}$: normal vector) be this optimal separation plane. and define the margin (${\sf marg}$) as the distance of the closest point ${\bf x}_{i}$ to $\cal{H}$, then it is easily seen that  ${\sf marg}=\frac{1}{\|{\bf w} \|}$. Thus, maximizing the margin lead to consider the following problem:
\begin{equation}
\left|
\begin{array}{l}
\medskip
\displaystyle{\min_{{\bf w},b}} \;\;\tau({\bf w}) \stackrel{\delta}{=} {\|{\bf w}\|}^{2}\;,\\
\mbox{s.t.\; :}\;y_{i}\left(\langle {\bf w},{\bf x}_{i} \rangle +b \right) \geq 1\;\;\forall \;i=1,\cdots, l\;\;y_{i}=\pm 1\;.
\end{array}
\right.
\label{svm2}
\end{equation}
Denoting $\Lambda$ the vector of Lagrange multipliers, dualization of eq. \ref{svm2} leads to consider again a quadratic problem, but with more explicit constraints \cite{cor:vap}, i.e. :
\begin{equation}
\left|
\begin{array}{l}
\medskip
\displaystyle{\max_{\Lambda}} \;W(\Lambda)=-\frac{1}{2}\;\Lambda^{T}\; D\; \Lambda+\Lambda^{T}\;\mathbf{1}\;,\\
\mbox{s.t. :}\;\Lambda \geq 0\;,\;\Lambda^{T} Y=0\;,
\end{array}
\right.
\label{svm3}
\end{equation}
where $\mathbf{1}$ is a vector made of $1$ and $Y^{T}=(y_{1},\cdots,y_{l})$ is the $l$-dimensional vector of labels, and $D$ is the Gram matrix:
\begin{equation}
D_{i,j}=\langle y_{i} {\bf x}_{i}, y_{j} {\bf x}_{j} \rangle \;.
\label{svm4}
\end{equation}
The dualized problem can be efficiently solved by classical quadratic programming methods. 
The less-perfect case consider the case when data cannot be separated without errors and lead to replace the constraints of eq. \ref{svm2} by the following ones:
\begin{equation}
y_{i}\left(\langle {\bf w},{\bf x}_{i} \rangle +b \right) \geq 1-\xi_{i}\;,\; \xi_{i}\geq 0\;,\ i=1,\cdots,l\;.
\end{equation}
Consider now a multiperiod extension of the previous analysis. Let us restrict first to a two-period analysis, we shall consider two separating hyperplanes (say ${\cal H}_{1}, {\cal H}_{2}$) defined by:
\begin{equation}
\left\{
\begin{array}{l}
\medskip
\langle {\bf w},x_{l}^{1}\rangle +b_{1} \gtrless \pm c_{1}\;\mbox{~according to:~}\;y_{l}^{1}=\pm 1\;,\\
\langle {\bf w},x_{l}^{2}\rangle +b_{2} \gtrless \pm c_{2}\;\mbox{~according to:}\;y_{l}^{2}=\pm 1\;.\\
\end{array}
\right.
\label{svm7}
\end{equation}
It is also assumed that these two separating planes are associated with time periods $T$ and $T+\Delta T$, $\Delta T$ known.
It is easily seen that the margin for the separating plane ${\cal H}_{1}$ is $\frac{c_{1}}{\|{\bf w}\|}$, while for the plane ${\cal H}_{2}$ it is $\frac{c_{2}}{\|{\bf w}\|}$. Thus, the problem we have to solve reads:
\begin{equation}
\left|
\begin{array}{l}
\medskip
\displaystyle{\min_{ {\bf w}, c_{1},c_{2},b_{1},b_{2}}} \quad \left[\displaystyle{\max_{1,2} }\;\left(\frac{ \|{\bf w}\|^{2}}{c_{1}^{2}},\frac{ \|{\bf w}\|^{2}}{c_{2}^{2}} \right)\;\right]\;,\\
\mbox{s.t.: }\;y_{l}^{1}\;\left(\langle {\bf w},x_{l}^{1}\rangle +b_{1}\right) \geq c_{1}\;,\;y_{l}^{2}\;\left(\langle {\bf w},x_{l}^{2}\rangle +b_{2}\right) \geq c_{2}\; \forall l.
\end{array}
\right.
\label{svm8}
\end{equation}
At a first glance, this problem appears as very complicated. But, without restricting generality, we can assume that $c_{1}<c_{2}$. This means that $
\displaystyle{\max_{1,2} }\;\left(\frac{ \|{\bf w}\|^{2}}{c_{1}^{2}},\frac{ \|{\bf w}\|^{2}}{c_{2}^{2}} \right)=\frac{ \|{\bf w}\|^{2}}{c_{1}^{2}}$. Making the changes $\frac{1}{c_{1}}\;{\bf w} \rightarrow {\bf w}^{'}$ and $\frac{b_{1}}{c_{1}} \rightarrow b_{1}^{'}$ then  leads to consider the classical problem:
\begin{equation}
\left|
\begin{array}{l}
\displaystyle{\min_{{\bf w}^{'},b_{1}^{'},b_{2}^{'}} } \;{\| {\bf w} \|}^{2}\;\\
\mbox{s.t. :} \;
\;y_{l}^{1}\;\left(\langle {\bf w}^{'},x_{l}^{1}\rangle +b_{1}^{'}\right) \geq 1\;,\;y_{l}^{2}\;\left(\langle {\bf w}^{'},x_{l}^{2}\rangle +b_{2}^{'}\right) \geq 1\; \forall l.
\end{array}
\right.
\label{svm9}
\end{equation}
Let ${\bf w}^{*}$ be the (unique) solution of eq. \ref{svm9}, then a straightforward calculation yields the distance $d({\cal H}_{1}^{*}, {\cal H}_{2}^{*})$ between the two separating planes, i.e.:
$$
d({\cal H}_{1}^{*}, {\cal H}_{2}^{*})=\frac{ |b^{*}_{1}-b^{*}_{2}|} { \| {\bf w}^{*}\|}\;.
$$
Finally, we deduce that the estimated velocity vector $\hat{\bf v}$ is given by:
\begin{equation}
\hat{\bf v}=\alpha\;{\bf w}^{*}\;\;\mbox{and:}\;\;\hat{v}=\frac{1}{\Delta T} \;d({\cal H}_{1}^{*}, {\cal H}_{2}^{*})\;.
\end{equation}
The previous analysis can be easily extended to an arbitrary number of periods, as long as the target trajectory remains rectilinear. Another definite advantage is that it can be easily extended to multitarget tracking.

\subsubsection{3D-SVM}

\begin{figure}[ht!]
{\includegraphics[width=9cm, height=8cm]{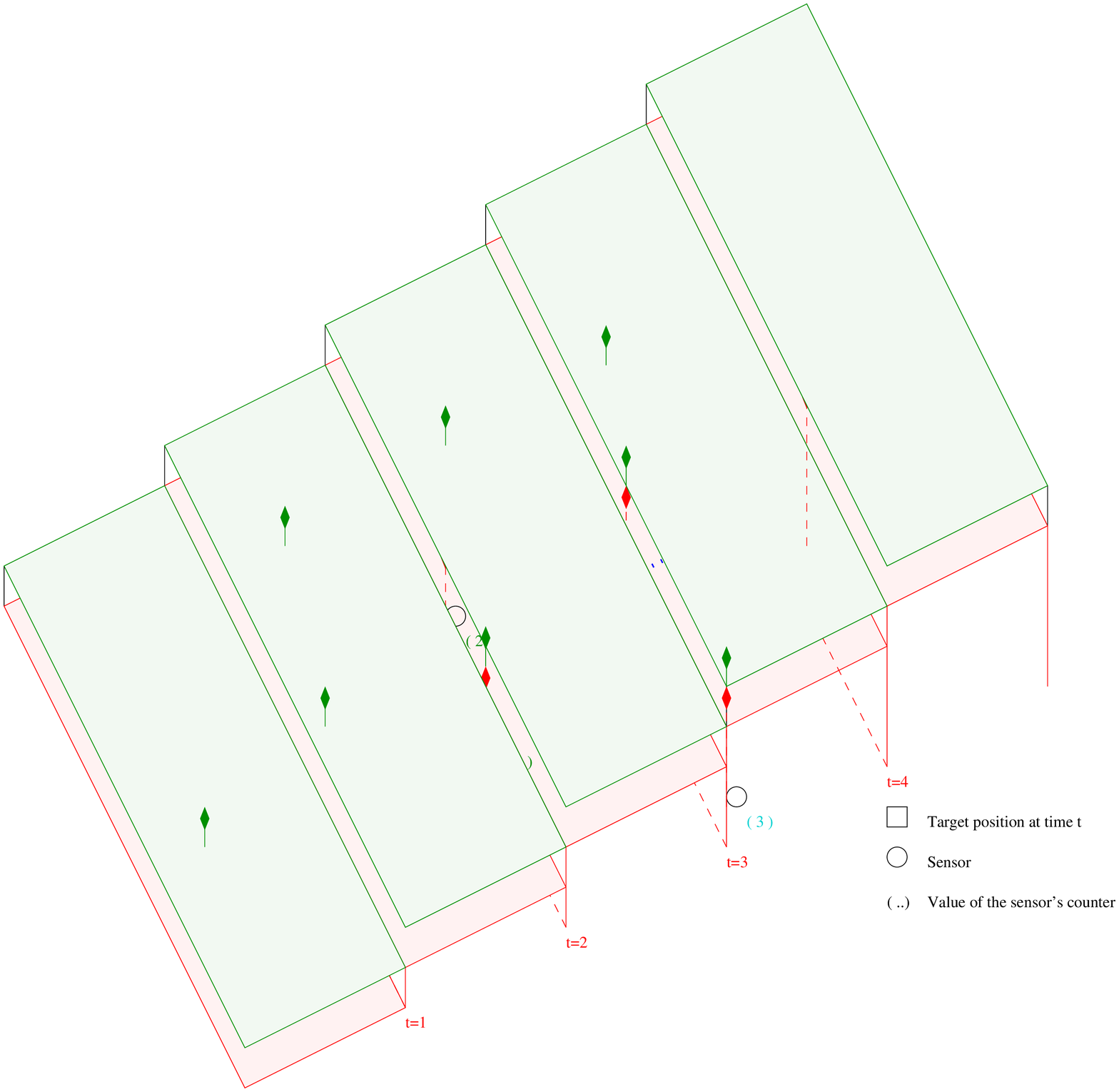}}
{\includegraphics[width=9cm, height=8cm]{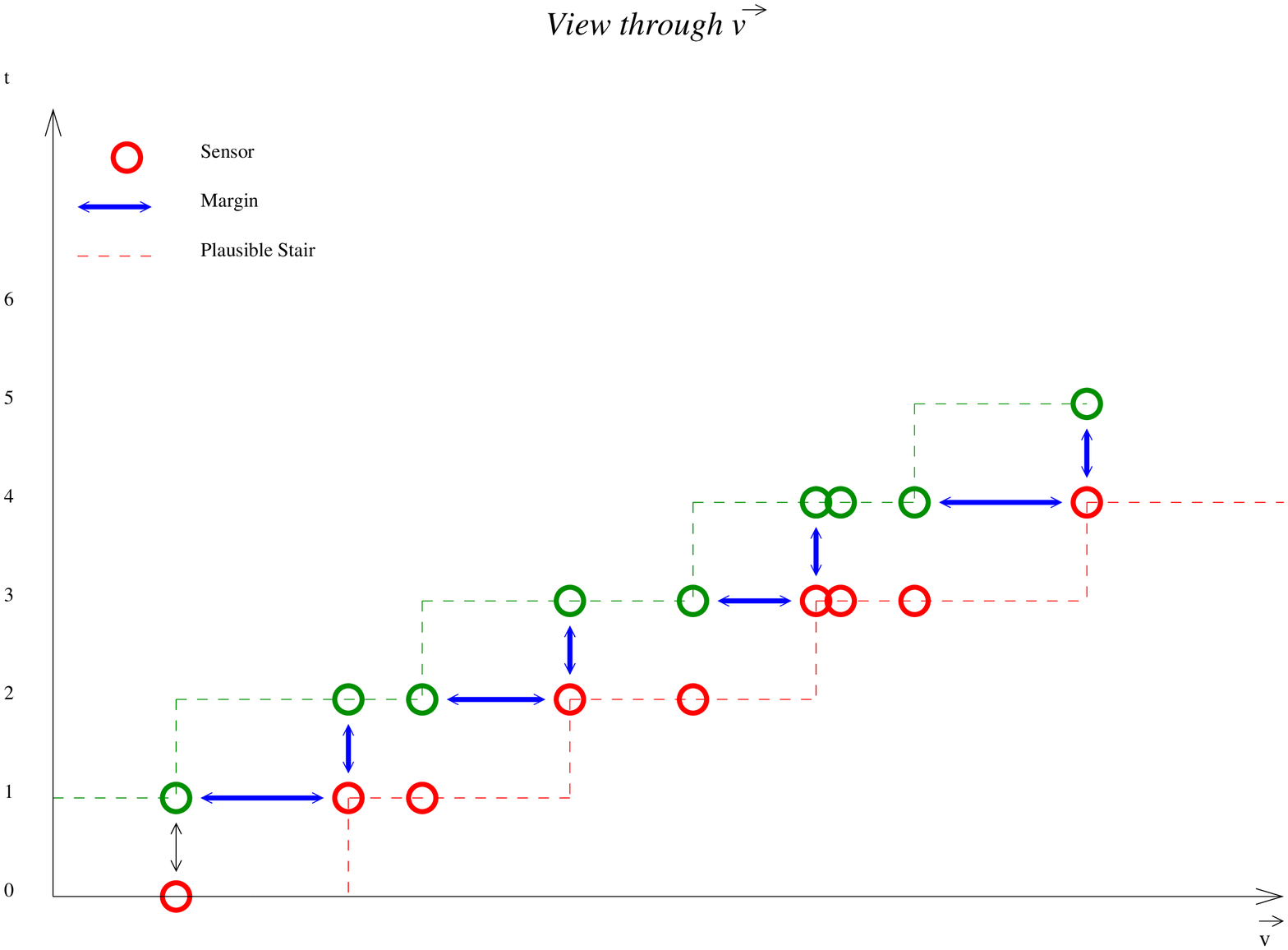}}
\caption{\it The theoretical stairway of the trajectory.}
\label{stair_svm}
\end{figure}

 We can also mix the SVM ideas with that of section \ref{stairwise}. Indeed, instead of focusing on a 2-D dataset, we can consider a 3-dimensional dataset (sensor coordinates and values of the sensor counters). The second 3-D dataset is the same, but the value of the counter is increased with {\it unity}. So, the separation plane is 2-D, and will be as closed  to the velocity plane as the sensor number can allow. See fig. \ref{stair_svm} for a more explicit understanding.
The results of the SVM estimation of the velocity plane are discussed in the Simulation Results section.

\subsection{Projection Pursuit Regression}

The projection pursuit methods have first been introduced by Friedman and Tuckey \cite{fri:tuc}. Then, they  have been developed for regression with the projection pursuit regression (PPR) by Friedman and Stuetzle \cite{fri:stu}. PPR is mainly a non-parametric method to estimate a regression, with however a certain particularity. Indeed, instead of estimating a function $f$ such as $Y_i = f(X_i) + \varepsilon_i$, where $X_i$ and $Y_i$ are known, and $\varepsilon_i$ assuming to follow a certain law, PPR estimates g such as $Y_i = g(X_i \;\mathbf{\theta}) + \varepsilon_i$. The first step of the algorithm is to estimate the direction $\mathbf{\theta}$, and then $\hat{g}$. In our specific case, $\mathbf{\theta}$ will represent the direction of the target, and $\hat{g}$ will give us the value of the velocity.

\subsubsection{Modeling }

Let $Y_i$ be the value of the $i$-th sensor counter. $X_i$ are the sensor coordinates. If $n(X_i \;\mathbf{\theta})$ is the value of the counter i at the end of the track, and $p$ the probability to have the right $\{+,-\}$ decision, we then have ($\mathcal{B}$: binomial):
\begin{eqnarray}
\mathcal{L} (Y_i \vert X_i \;\theta ) = \mathcal{B}(n(X_i \;\theta), p)
\end{eqnarray}
Assuming in a first time that $p=1$, the two parameters we would like to estimate are the $\mathbf{\theta}$ parameter and the $n(.)$ function.

\subsubsection{The PPR method in the network context}
We have some additional constraints on $n(.)$. First of all, it only takes integer values. Then, it is an increasing function (because $p=1$). The optimization problem we have to solve is the following:
\begin{eqnarray}
\hat{\theta} &=& \displaystyle{\arg \min_{\mathbf{\theta}} }\;\sum\;(\hat{n}(X_i \;\theta) - Y_i)^2\;,
\end{eqnarray}
where $\hat{n}$ is calculated in a quite special way. First, we define a non parametric estimation of a function f, via:
\begin{eqnarray}
\hat{f}(u) &=& \frac{\sum{Y_i K_h(X_i \;\theta - u)}}{\sum{K_h(X_i \;\theta - u)}}\;.
\end{eqnarray}
Then, we sort $(X \; \mathbf{\theta})_i$ into a vector $(X \theta)_{(i)}$ from the smallest to the biggest. After which we define $\hat{n}(.)$ via:
\begin{equation}
\left\{
\begin{array}{lll}
\hat{n}(X \;\mathbf{\theta}_{(i)}) &=& \hat{f}(X \;\mathbf{\theta}_{(i)}) \quad \textrm{if} \quad \hat{f}(X \;  \mathbf{\theta}_{(i)}) \geq \hat{f}(X \;\mathbf{\theta}_{(i-1)}) \;, \\
\hat{n}(X\; \mathbf{\theta}_{(i)} ) &=& \hat{f}(X \; \mathbf{\theta}_{(i-1)}) \quad \textrm{otherwise}\;.
\end{array}
\right .
\end{equation}
Sometimes, due to the integer value of the estimated $n(.)$ function, we have to deal with many possible values of  $\widehat{\mathbf{\theta}}$. Then, in this case, we choose the mean value of $\mathbf{\theta}$. Due to the specific behavior of our target and our modeling, we know in addition that the general form of n (say $\tilde{n}$) is given by:
\begin{eqnarray}
\tilde{n}(u) &=& \sum i \mathbb{I}_{[(X \mathbf{\theta})^{\perp} + (i-1)v, (X \mathbf{\theta})^{\perp} + i v]}(u)\;.
\end{eqnarray}
The next step is then to estimate $v$. Such an estimation is given by the following optimization program:
\begin{eqnarray}
\hat{v} &=& \displaystyle{\arg \min_{v}} \sum(\hat{\tilde{n}}(X \widehat{\mathbf{\theta}_i }) - Y_i)^2
\end{eqnarray}
\subsubsection{Convergence}
We will study if the estimation is good with an infinite number $N$ of sensors. Assuming we have an infinite number of sensors in a closed space, this means that each point of the space gives us an information $\{+,-\}$. We then will have the exact parameters of the stairwise functional.
To that aim, we will show in the following paragraph that the probability of having a sensor arbitrary close to the limits of each  stair steps is $1$.
We assume that the sensor  positions are randomly distributed, following an uniform law. Then, $y$ being fixed:
\begin{eqnarray}
\mathcal{L}(X \vert Z) &=& \mathcal{U}_{[B_{inf}; B_{sup}]}
\end{eqnarray} 
If the velocity vector $\bf{v}$ is denoted with $[a;b]$, then:
\begin{equation}
\begin{array}{ccccc}
B_{\inf} &=& -\frac{b}{a} y - \frac{c_{\inf}}{a} \;, \qquad B_{\sup} &=& -\frac{b}{a} y - \frac{c_{\sup}}{a}\;,
\end{array}
\end{equation}
where $(c_{\inf},c_{\sup})$ only depends on $\bf{v}$ and ${\bf x}_{0}$, which means that they are deterministic, and independent from $X$. It is quite obvious that $B_{\inf}$ represents the smaller $x$-limit of a step, when $B_{\sup}$ represents its higher $x$-limit. Then, considering the velocity plane, $B_{\inf}$ and $B_{\sup}$ both belong to the plane. Denote $u = \inf_i (X_i)$, then:
\begin{eqnarray}
\forall \varepsilon>0 \qquad P(\vert u - B_{\inf} \vert < \varepsilon) &=& P( u-B_{\inf} < \varepsilon) \;,\\ \nonumber
{}&=& P( u < \varepsilon + B_{\inf})\;.
\end{eqnarray}
where we note $\mathcal{A} = \vert u - B_{\inf} \vert < \varepsilon$. We know that:
\begin{equation}
P(\inf X_i \leq t) = \left \{
\begin{array}{ll}
0 & \textrm{if $t \leq B_{\inf}$}\;,\\
1 - (\frac{B_{\sup}-t}{B_{\sup}-B_{\inf}})^N & \textrm{if $t \in [B_{\inf};B_{\sup}]$} \;,\\
1 & \textrm{if $t > B_{\sup}$} \;.
\end{array}
\right.
\end{equation}
Then, we have the following probability calculations:
\begin{eqnarray*}
P(\mathcal{A}) &=& P(\vert u - B_{\inf} \vert < \varepsilon)\;,\\
{} &=& P( u < \varepsilon + B_{\inf}) \;, \nonumber\\
{} &=& 1 - (\frac{B_{\sup}-(\epsilon + B_{\inf})}{B_{\sup}-B_{\inf}})^N 1_{[B_{\inf};B_{\sup}]}((\varepsilon + B_{\inf})) \;, \nonumber\\
{} &=& \left \{ 
\begin{array}{ll}
0 & \textrm{if $\varepsilon \leq 0$}\\
1 - (1-\frac{\varepsilon}{B_{\sup}-B_{\inf}})^N & \textrm{if $\epsilon \in ]0;B_{\sup}-B_{\inf}]$} \\
1 & \textrm{if $\varepsilon > B_{\sup}-B_{\inf}$} \;.
\end{array}
\right.
\end{eqnarray*}
Given the above equation, $1-\frac{\varepsilon}{B_{\sup}-B_{\inf}}$ is smaller than one, which means that $(1-\frac{\epsilon}{B_{sup}-B_{inf}})^N$ converges to $0$ as $N$ increases to infinity. Thus, we have finally:
\begin{eqnarray}
\forall \varepsilon>0 \qquad \lim_{N \to \infty} \qquad P(\vert u - B_{\inf} \vert < \varepsilon) &=& 1\;.
\end{eqnarray}
ending the proof.

\section{Non-linear trajectory estimation}
\label{obs}

\subsection{Target Motion Model}

The target is assumed to evolve with a Markov motion, given by:
\begin{eqnarray}
\bf{x_k} \vert \bf{x_{k-1}} & \sim & \mathcal{N}(F_k \bf{x_{k-1}}, Q_k) 
\end{eqnarray}
for $k=1,2...$ where $\mathcal{N}(\mu, \sigma^2)$ is a gaussian distribution with mean $\mu$ and variance $\sigma^2$. The starting position is assumed to be unknown.

\subsection{Sensor Measurement Model and Analysis}

At each time period, each sensor gives us a $\{+,-\}$ information, meaning that the target is getting closer or moving away. Given all the sensors reports at the time-period $t$, we can easily define a space where the target is assumed to be at this time-period. This is the fundamental uncertainty we have at $t$, and the area of this domain is, of course, directly related to the network parameters (sensor number, network geometry, etc.). 

\subsection{Velocity Estimation}

We can estimate the direction of the target based on the simple information given by the sensors. Obviously, that estimator will only be precise if the number of sensors is significantly great. To perform that estimation, we can use several methods, such as the Projection Pursuit Regression Method, or the Support Vector Machine Method. The SVM method chosen for our algorithm as a most common method, and is presented in the next paragraphs.

\subsubsection{The effect of target acceleration}

To illustrate the effect of velocity change for estimating the target position, let us consider a very simple example. Assume that the target motion is uniformly accelerated, i.e. :
\begin{equation}
{\bf x}_{t}={\bf x}_{0}+t \;{\dot{\bf x}}_{0}+ t^{2} \;{\ddot{\bf x}}_{0}\;.
\label{acc1}
\end{equation}
We have now to deal with the following question: Is the target trajectory fully observable? To that aim, we first recall the following result. Considering a dense binary network, two target trajectories are said indistinguishable iff they provide the same (binary) information which is equivalent to the following conditions:
\begin{equation}
\left\{
\begin{array}{l}
{\dot{\bf x}}_{t}={\dot{\bf y}}_{t} \;,
\langle {\bf y}_{t} -{\bf x}_{t}, {\dot{\bf y}}_{t} \rangle =0\;\; \forall t\;.
\end{array}
\right.
\label{acc2}
\end{equation}
Expliciting the second condition of eq \ref{acc2}, with the target motion model \ref{acc1}, we obtain that the following condition holds ($\forall t$):
\begin{equation}
\begin{array}{l}
\langle {\bf y}_{0} -{\bf x}_{0}, {\dot{\bf y}}_{0} \rangle + t \langle {\dot{\bf y}}_{0}- \dot{\bf x}_{0}, {\dot{\bf y}}_{0} \rangle +\frac{1}{2} t^{2}\langle {\ddot{\bf y}}_{0}- {\ddot{\bf x}}_{0}, {\dot{\bf y}}_{0} \rangle\;,\\
+ t\langle {\bf y}_{0} -{\bf x}_{0}, {\ddot{\bf y}}_{0} \rangle +t^{2}\langle {\dot{\bf y}}_{0}-\dot{\bf x}_{0}, {\ddot{\bf y}}_{0} \rangle+\frac{1}{2} t^{3}\langle {\ddot{\bf y}}_{0}- {\ddot{\bf x}}_{0}, {\ddot{\bf y}}_{0} \rangle=0\;.
\end{array}
\label{acc3}
\end{equation}
Thus, $\langle {\bf y}_{t} -{\bf x}_{t}, {\dot{\bf x}}_{t} \rangle$ is a zero polynomial, which means that all its coefficients are zero. For the $t^{3}$ coefficients we obtain the condition $\langle {\ddot{\bf y}}_{0}- {\ddot{\bf x}}_{0}, {\ddot{\bf y}}_{0} \rangle=0$. Similarly with the $\langle {\bf y}_{t} -{\bf x}_{t}, {\dot{\bf x}}_{t} \rangle =0$, we obtain $\langle {\ddot{\bf y}}_{0}- {\ddot{\bf x}}_{0}, {\ddot{\bf x}}_{0} \rangle=0$. Subtracting these two equalities yield $\| {\ddot{\bf y}}_{0} -{\ddot{\bf x}}_{0} \|=0$ , or ${\ddot{\bf x}}_{0}={\ddot{\bf y}}_{0}$.\\

Quite similarly, we obtain the equality ${\dot{\bf x}}_{0}={\dot{\bf y}}_{0}$ and the last equality:
\begin{equation}
\langle {\bf y}_{0} -{\bf x}_{0}, {\dot{\bf y}}_{0}+ t {\ddot{\bf y}}_{0} \rangle =0 \;\; \forall t\;.
\end{equation}
Assuming that the couple $\left\{  {\dot{\bf y}}_{0}, {\ddot{\bf y}}_{0} \right\}$ spans the sensor space then we deduce that ${\bf x}_{0}={\bf y}_{0}$. So, it has been shown that it was the target acceleration which render the problem fully observable. This reasoning can be extended to a wide variety of target modeling.

\subsection{Tracking algorithm}

\begin{figure}[ht!]
\center
{\includegraphics[width=9cm, height=6cm]{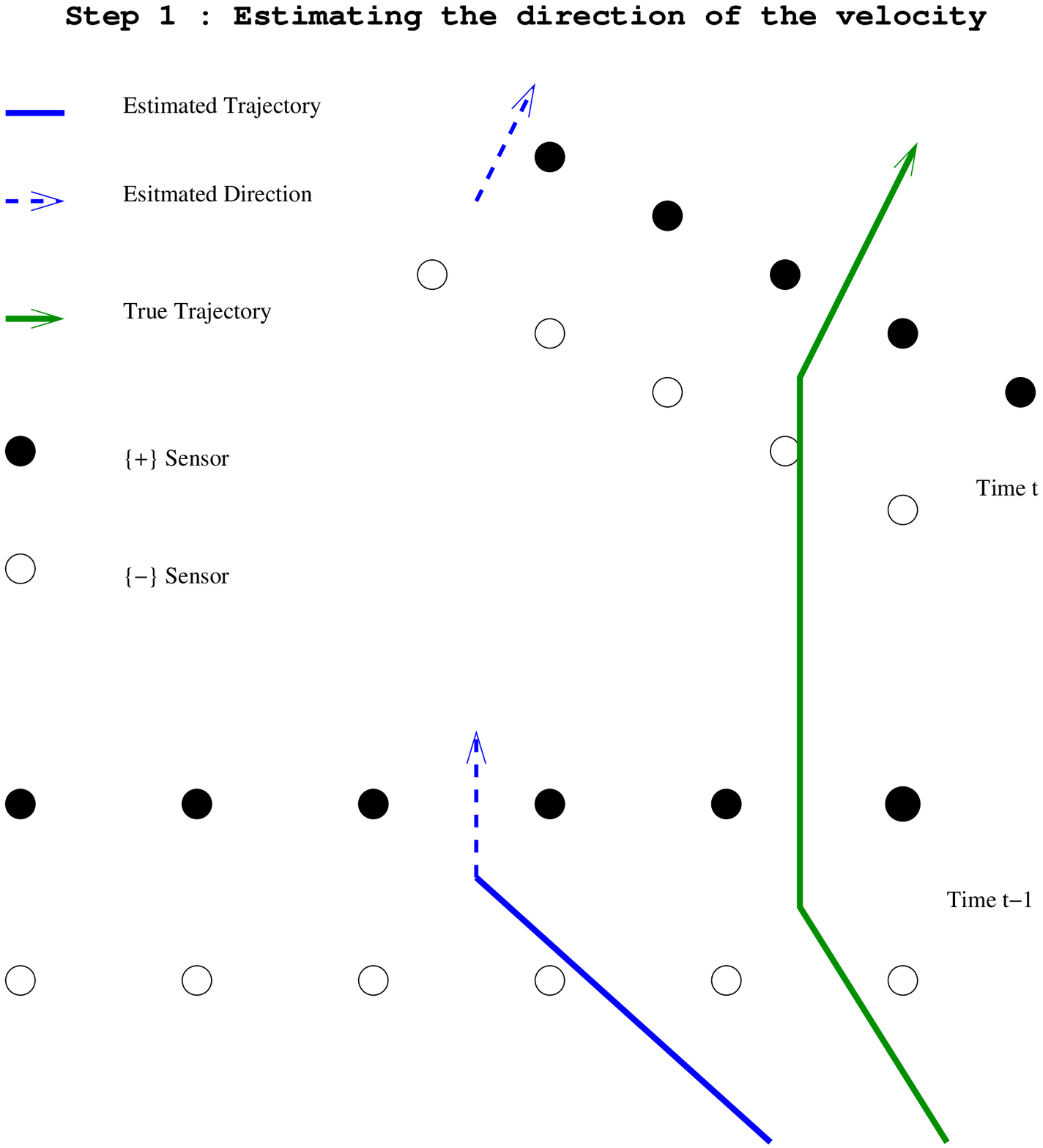}}
{\includegraphics[width=9cm, height=6cm]{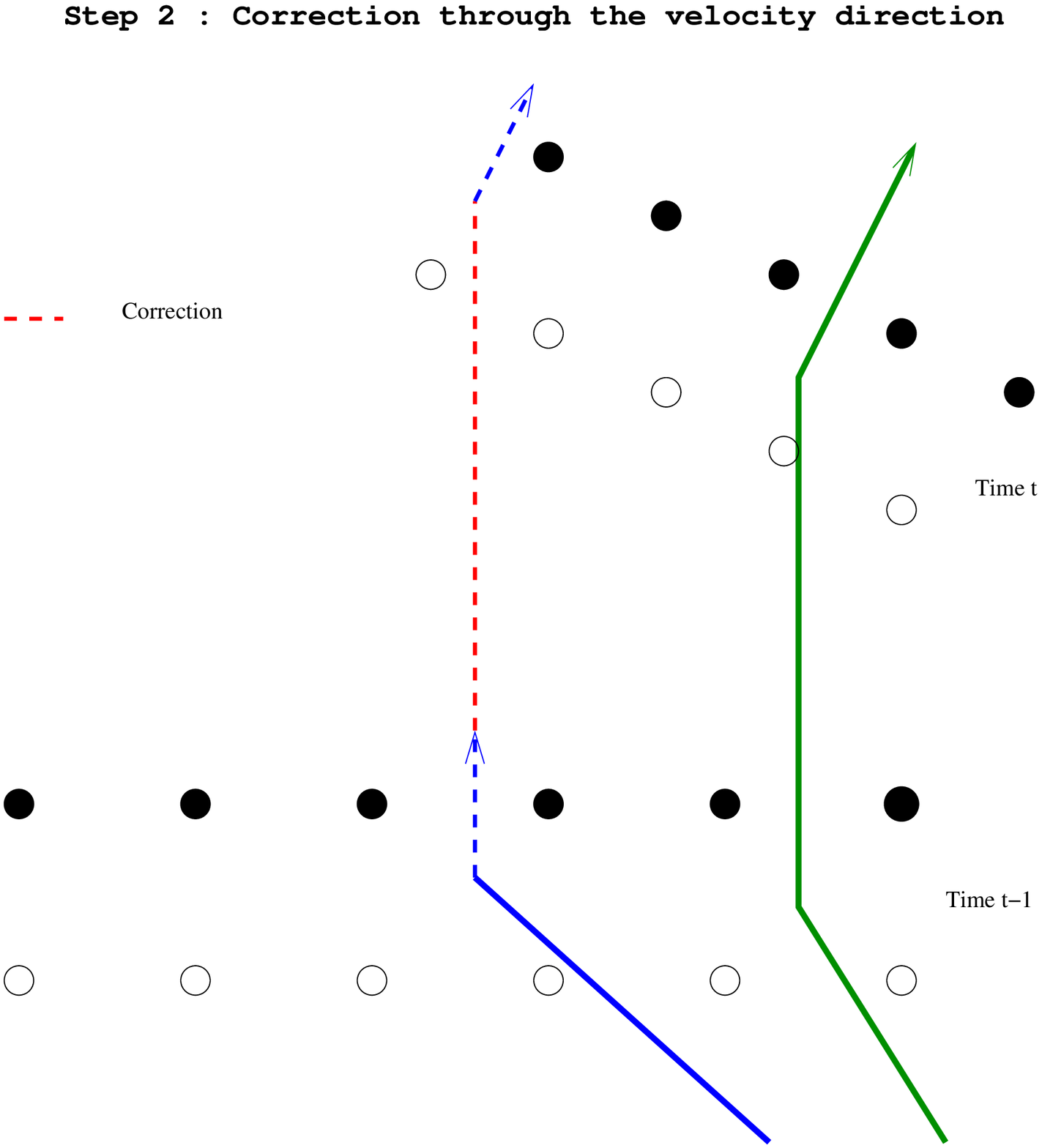}}
{\includegraphics[width=9cm, height=6cm]{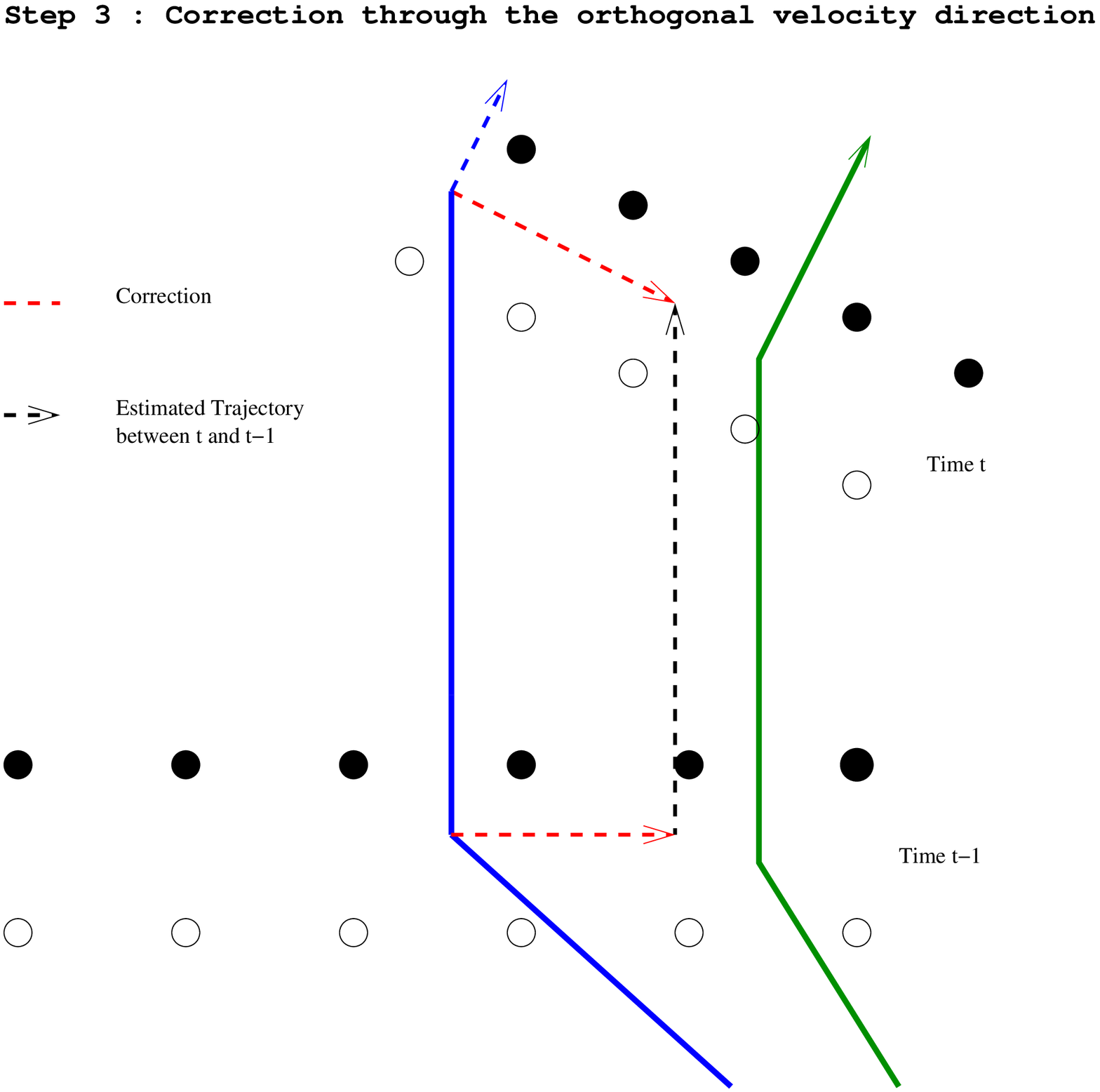}}
\caption{\it Correction scenario.}
\label{schema_correction}
\end{figure}

The main issue with the SVM estimation is that it only provides us the general direction of the target within a deterministic framework. Moreover, it is highly desirable to develop a reliable algorithm for target tracking (velocity and position). To solve this problem, we build a two-step algorithm. In the first step, we perform a correction through the estimated unitary velocity vector at each time-period $t$, called $\lambda_t$. Then, in a second time, we perform a correction through the orthogonal-estimated (unitary) velocity vector, also at each time-period, called $\theta_t$. These two corrections give us a better estimation of both the velocity and the position of the target. We refer to fig. \ref{schema_correction} for the presentation of the rationale of the two correction factors. 

\subsubsection{The $\lambda$ factor}

To build that correction factor, we started with a very simple assumption. At each period $t$, the sensors provide  binary motion information. Thanks to the first part of this article, we know that the target is in the (special) set lying  between the two same-sign-sensors set. Then, starting from the previous estimated position of the target, we move the estimated target through the estimated velocity vector direction until it stands in that special set. We now define this operator in a mathematical way:\\
Let ${\hat{\bf v} }_{t}$ the estimated normalized velocity vector at time $t$.\\
Moreover, let $\{ {\bf t}_i^{(-)}\}_i$ (respectively $\{ {\bf t}_i^{(+)}\}_i $) the coordinates of the sensors ($s_{i}$) giving a $\{-\}$ (respectively a $\{+\}$) at time $t$.\\
We sort $vs_i^{(-)} = \langle {\hat{\bf v} }_{t} ,{\bf t}_i^{(-)} \rangle$ (respectively $vs_i^{(+)}= \langle {\hat{\bf v} }_{t} ,{\bf t}_i^{(+)} \rangle$). Then, following a very simple geometrical reasoning, we note that $\langle \hat{\bf{v}_t}; \hat{X_t} \rangle$ should be between $vs_{max}^{(-)}$ and $vs_{min}^{(+)}$. To ensure that property, we define the following correction factor:
\begin{equation}
\begin{array}{l}
\lambda_{t} = \frac{ vs_{moy}^{(+,-)} - \langle {\hat{\bf v} }_{t}, {\hat{\bf x}}_{t-1} \rangle}{\langle {\hat{\bf v} }_{t},  {\hat{\bf v} }_{t-1}\rangle}\;,\\
\mbox{with the following definition of } vs_{moy}^{(+,-)}\; :\\
vs_{moy}^{(+,-)}= \frac{vs_{max}^{(-)}+ vs_{min}^{(+)}}{2} \;,
\end{array}
\label{correct1}
\end{equation}
To calculate this factor, we consider the projection equality:
\begin{equation}
\langle  {\hat{\bf v} }_{t}, ( {\hat{\bf x}}_{t-1} + \lambda_t \;{\hat{\bf v} }_{t-1}) \rangle= vs_{moy}^{(+,-)}\;
\label{correct2}
\end{equation}
which means that the projection of the corrected value is equal to the mean value of the projection. Geometrically, this means that the position of the target is estimated to be in the center of the special set defined by the sensors. The value of the correction factor $\lambda_{t}$ (see eq. \ref{correct1}) is then straightforwardly deduced from eq. \ref{correct2}. Similarly, the target position is updated via:
\begin{equation}
{\hat{\bf x}}_{t}^{corr} = {\hat{\bf x}}_{t-1} + \lambda_t \;{\hat{\bf v} }_{t-1}\;.
\end{equation}
Here the correction factor $\lambda_{t}$ has been calculated via the average value of the projection. This is an arbitrary choice and we can consider the lower or the upper bound  of the projection with no significant difference on the results of the algorithm.\\
Obviously, if  the estimation of the position is not very good, the estimated velocity value (clearly based on $\lambda_t$) will be quite different from the real value of the velocity. The next correction factor is based on the assumption that the target velocity changes are upper and lower bounded.
  
\subsubsection{The $\theta$ correction factor}

We assume that the velocity of the target has bounded acceleration. Then, if the velocity estimated at a certain time $t$ is too different from the velocity estimated at time $t-1$, this means that the estimated position of the target is far from the right one. Then, in that precise case, we consider an orthogonal correction, through $\hat{\bf{v}_t}^{\perp}$. \\
For that deterministic algorithm we decided to perform a very simple modeling of the velocity. Indeed, we take as a right value for the velocity the simple mean of the $k$ previous values of the estimated velocity ($m_{t,k}$). We calculate in addition the variance ($\sigma_{t,k}$), and the factor $\theta_t$ can be non-zero iff the estimated value of the velocity at time $t$ is not in the interval given by  $[m_{t,k} - \sigma_{t,k}; m_{t,k} + \sigma_{t,k}]$. We then look for $\theta_t$ such that:
\begin{equation}
\langle {\hat{\bf x}}_{t}^{corr}+ \theta_t\; {\hat{\bf v} }_{t}^{\perp} - ( {\hat{\bf x}}_{t-1}+ \theta_{t} \;{\hat{\bf v} }_{t-1}^{\perp} ) ;  {\hat{\bf v} }_{t-1} \rangle = m_{t,k} \;.
\end{equation}

The previous equation needs some explanation. Given that ${\hat{\bf x}}_{t}$ is the estimated  target position at time $t$, we would like to correct the value to be closer to the right position. The only way we can deal with it, is to correct the estimated value of the velocity. ${\hat{\bf x}}_{t}^{corr}- {\hat{\bf x}}_{t-1}$ is the previous calculated correction. If the difference between that estimation and the value $m_{t,k}$ is too important, we try to reduce that difference with a translation of the positions at time periods $t$ and $t-1$. As we want the positions to stay in the special set defined by the sensors, the direction of that translation is given by ${\hat{\bf v} }_{t}^{\perp}$ for the position at time $t$, and $\hat{\bf{v}}^{\perp}_{t-1}$ for the position at time $t-1$.\\
Performing straightforward calculation, leads to consider the following correction factor:
\begin{equation}
\theta_t = \frac{ m_{t,k}- \lambda_t}{ {\langle \hat{\bf{v}}_t^{\perp} ; \hat{\bf{v}}_{t-1} \rangle} }\;.
\end{equation}

Obviously, as we could expect when presenting the method, if the target motion is rectilinear and uniform , no correction factor can be calculated.
Then, the final estimated position is given by:
\begin{equation}
{\hat{\bf x}}_{t}^{fin} = {\hat{\bf x}}_{t}^{corr} + \theta_t\; \hat{\bf{v}}_t^{\perp} \;.
\end{equation}

\subsubsection{The final correction step}

Noticeably the most important step of the algorithm, i.e. the $\theta$ correction factor, is based on the estimation of the  velocity change. Indeed, the best the estimation of the velocity is, the best we can estimate the position. Then, our aim is to perform a better analysis of the target motion.
Considering that from time to time, the estimation of the position increases in quality, a promising way should be to perform  a feedback of the newest corrector to the oldest position estimation.
We denote ${\hat{\bf z}}_{t}$ the updated estimated position of the target at time $t$. Then, according to the previous paragraph, the estimated position is updated via:
\begin{equation}
\forall j<t: \;\:{\hat{\bf z}}_{j} = {\hat{\bf x}}_{j}^{fin} + \sum^{t}_{i=j+1} \theta_i \hat{\bf{v}}_i^{\perp}\;.
\end{equation}

With this new estimator we will be able to perform a better analysis of the target motion (position and velocity).

\subsubsection{The final algorithm}

With the definition of the correction factors, the theoretical part of the algorithm is finished. Then, it is presented as follows, at time period $t$:\\
\begin{enumerate}
\item Get the binary information of each sensor, and then the target position set.\\
\item Estimate the velocity direction at time $t$ via a SVM method\\
\item Perform the $\lambda$ calculation, and add that correction to the estimated velocity at time $t-1$. The time-$t$-position is then updated.\\
\item Check if the estimated velocity at time $t-1$ is too different from the modeled value, and in this case, calculate $\theta_t$.\\
\item Update the position at time $t$, and in this case, the velocity at time $t-1$ with the correction $\theta$.  
\end{enumerate}
Step 2 and 3 can be inverted with no damage in the process.
This is the main part of the algorithm. However, there is no mention in that enumeration of the initialization. There are two main state vectors that have to be initialized. The position and the velocity. The position is assumed to be unknown, but thanks to the sensors, we can have a space where the target is assumed to be at first. We use here a uniform law for the initialization, given that we have no further information about where the target can start.\\
The initialization of the velocity is not far from that solution. Indeed, with the binary information, we can provide a convenient estimate of the velocity direction. Even if we don't have a precise idea of the speed value, we can then start the algorithm.

\section{Simulation Results}

\subsection{Constant Velocity movement}

\begin{figure}[ht!]
{\includegraphics[width=9cm, height=8cm]{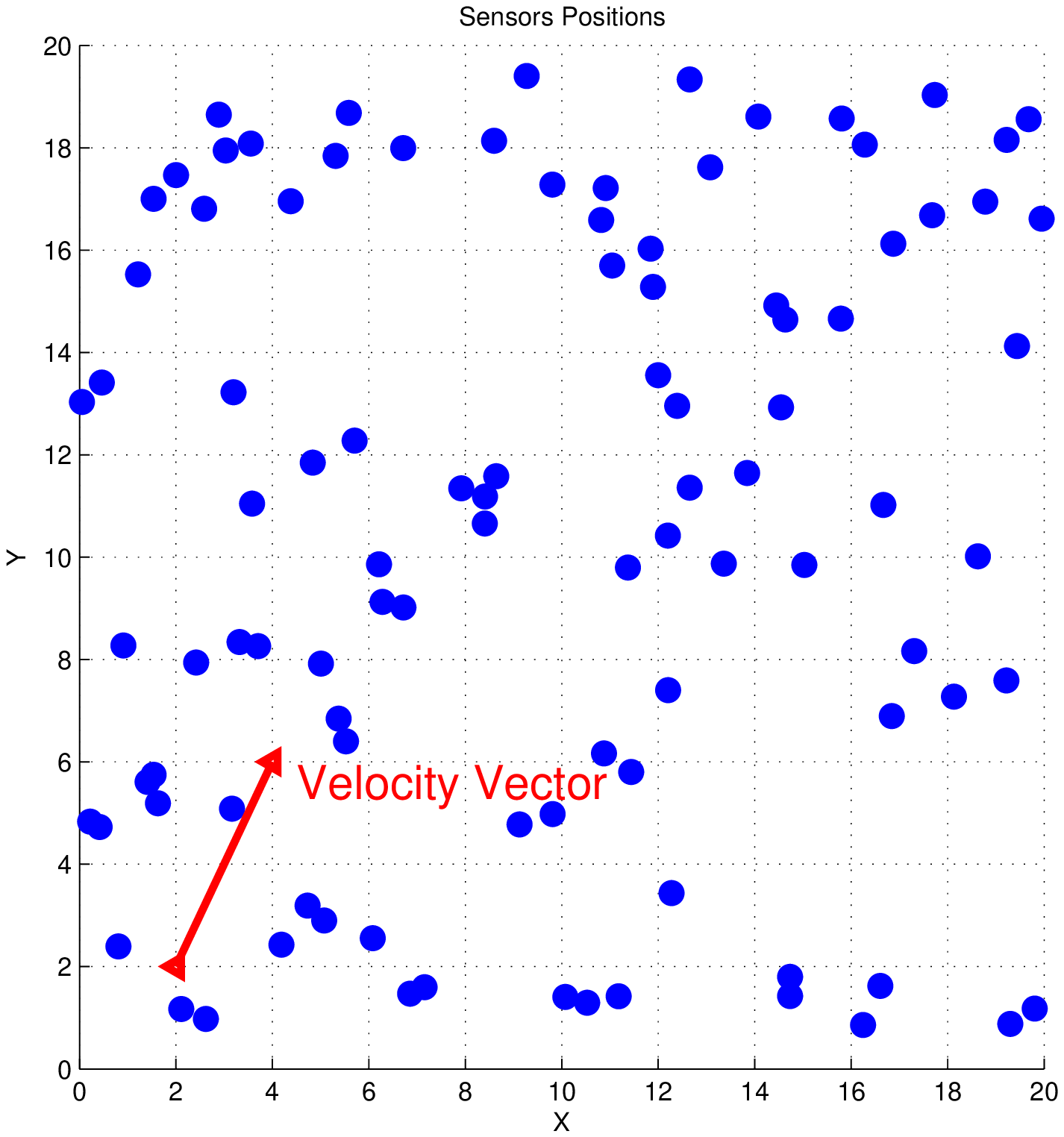}}
{\includegraphics[width=9cm, height=8cm]{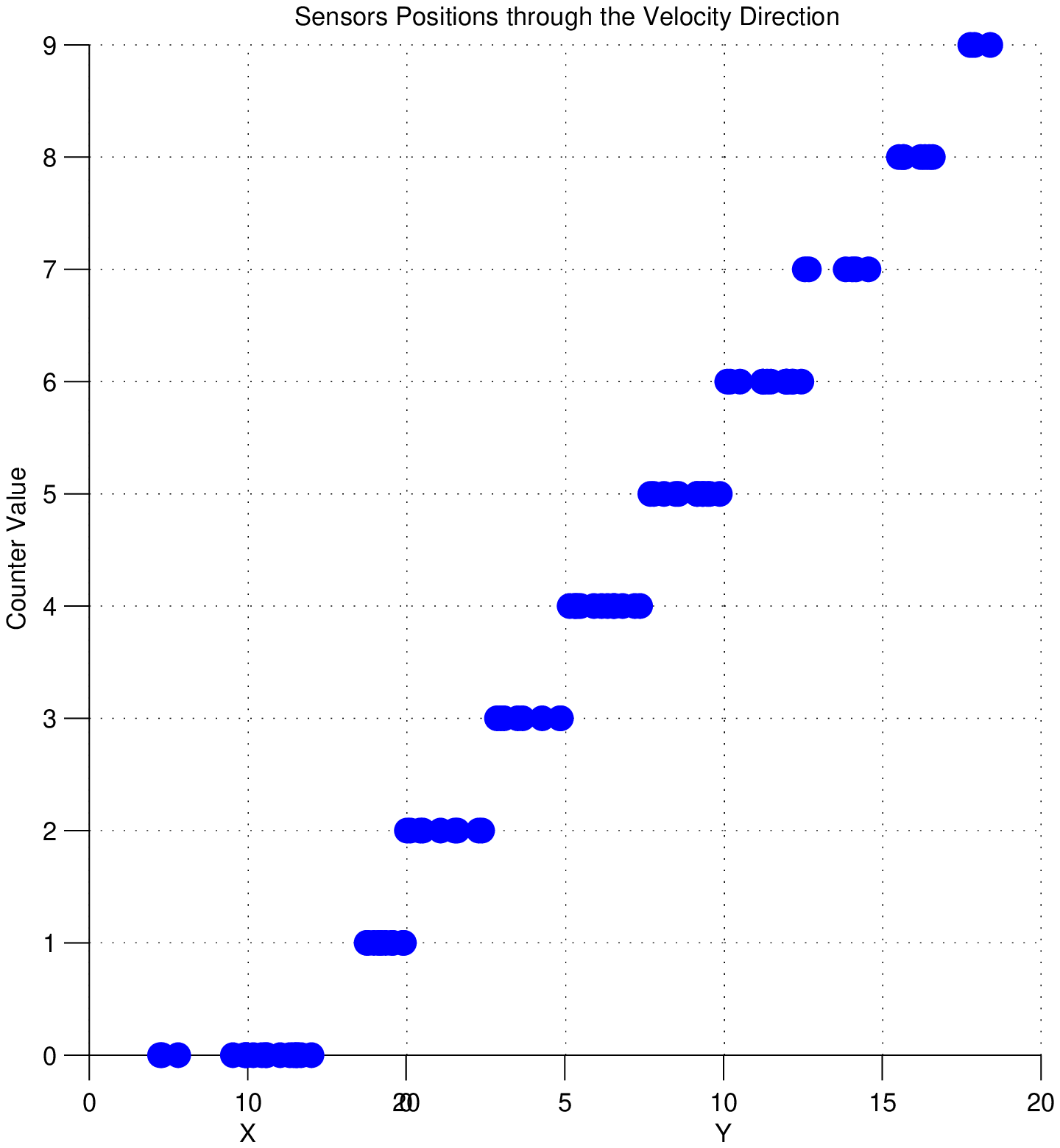}}
\caption{\it 2D sensor position and velocity's direction projection. N=100 sensors, Velocity Vector is [1,2] }
\label{simulation_stair}
\end{figure}

\begin{figure}[ht!]
\begin{center}
{\includegraphics[width=20cm, height=8cm]{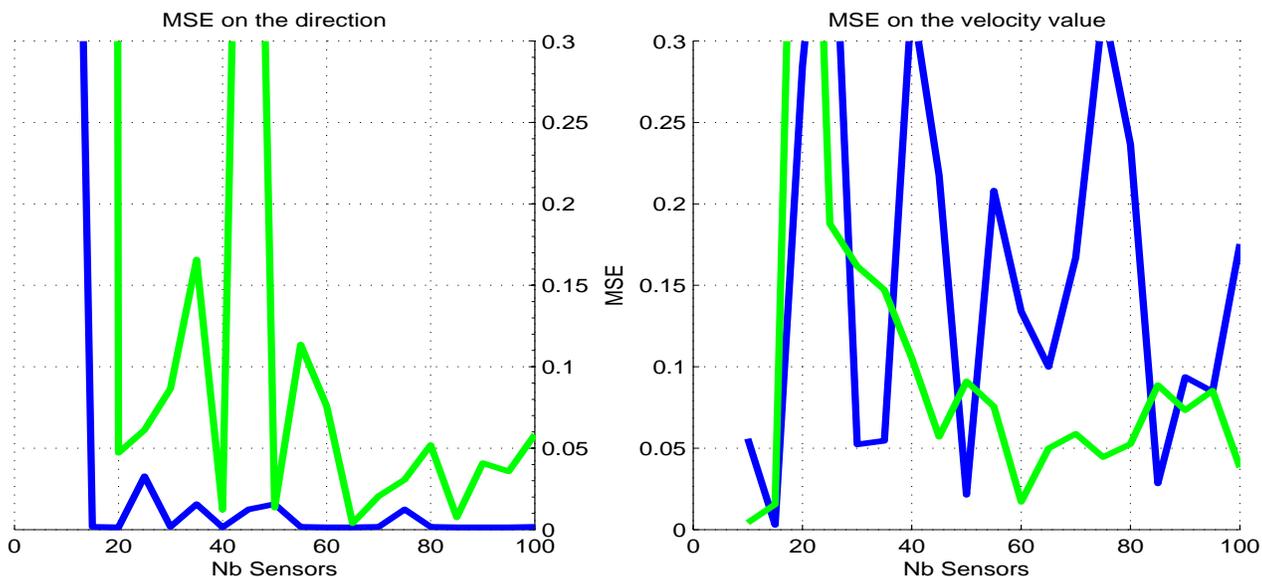}}
\caption{\it Mean Square Error of the Velocity Estimators. Green for the SVM, Blue for the PPR.}
\label{sim_conv}
\end{center}
\end{figure}

We shall now investigate the previous developments via simulations. The first figure (fig.\ref{simulation_stair}) will show the stair built by the previously explained method ($N$: fixed). The position of the sensors are considered random, following a uniform law on the surveillance set.
To evaluate the performance of our methods, we decided to calculate the mean square error of the two estimated parameters, which are the velocity value and the velocity direction. Fig. \ref{sim_conv} shows the two MSEs values for both direction and velocity values, assuming the sensor number is growing from $10$ to $100$, and the velocity vector is the $[1,2]$ vector ($m/s$).\\
Providing 2000 simulations, the MSEs seems to be unstable. However, the two parameter estimation methods leads to a very different conclusion. In the case of the direction estimation, the PPR method works highly better than the SVM method, and seems quite stable as the sensor number $N$  grows. On the other side, the SVM method is more erratic.One possible explanation is that the PPR method has been first developed for the particular case of direction estimation, while the SVM method is more focused on the margins maximization, which means in our case a simultaneous estimation of both parameters.
The conclusions we can make on the velocity value estimation are rather opposite. The MSE becomes reasonable only for the SVM method, and for a number of sensors up to 60. Indeed, we have a $0.05$ $m/s$ error on a velocity value estimation for a theoretical value of $\sqrt{5}$. As erratic as the SVM's MSE was in the direction estimation, it was however less erratic than the result we have for the PPR value. \\
One answer to the MSE erratic value for the PPR could be to find a best way to estimate the velocity value. Indeed, in our case, we choose for estimating functional a sum of indicators functions. However, it is not clear that this optimization gives a single minimum solution. There could be a finest functional that could lead to a most robust optimization solution, and  this would be the subject of future works.

\subsection{Random walk}

\begin{figure}[ht!]
\center
{\includegraphics[width=9cm, height=8cm]{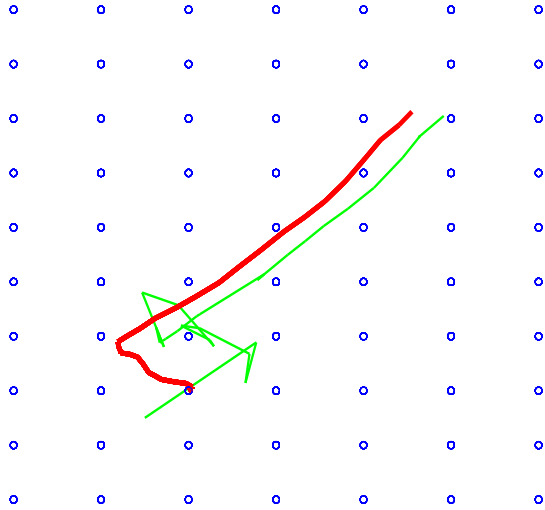}}
\caption{\it Trajectory Estimation of a target. In Red, the real trajectory, in green the estimated one.}
\label{traj_est}
\end{figure}

We will present in that section the results of the tracking algorithm. We consider here that the target starts from the $[100,100]$ position and that its initial velocity vector is the $[1,1]$ vector. The number of sensors is equal to $70$, in a quite wide space (300mx300m). The variance of the target motion  is not very important, and the tracking duration is $T=30$ seconds.\\
One simulation is presented  in figure \ref{traj_est}. In red is represented the real target trajectory, quite diffusive, and in green the estimated successive positions. The initialization is not very bad because the number of sensors is quite important, which means the uniform set is not too large. After the first step, the estimation seems to hang the real trajectory, and follows the target well (less than $10$ meter error). However, when the target turns right, we loose some precision, mainly because the correction factors seemed to be ``lost''. The reason for that behavior is that the SVM method provides us a bad estimation of the velocity vector. Then, the algorithm provides a correction in a bad direction, which moves away from the real trajectory. During a few seconds, the estimation works quite bad, before hanging again the target direction, and then performing a quite good estimation of the velocity. Unfortunately, there is no evidence in that example that increasing indefinitely the tracking duration results in an  estimated position closer and closer to the real target position.
This is precisely the aim of the two next figures in \ref{mse_state}. The first one shows the mean square error of the estimated position of the target through the trajectory. The total time is $T=30$ seconds, and we can see an amazing  and remarkable decrease of that MSE in the first seconds. It seems however that there is a limit to that decrease. Indeed, the MSE will not converge to a zero value, even if we could perform a long-time tracking. Clearly, the limitation is due to the binary information at first, and certainly to the number of sensors in a second time. Some further work could certainly exhibits a strong link between the number of sensors and the MSE of the position.\\ 
\begin{figure}[ht!]
{\includegraphics[width=9cm, height=8cm]{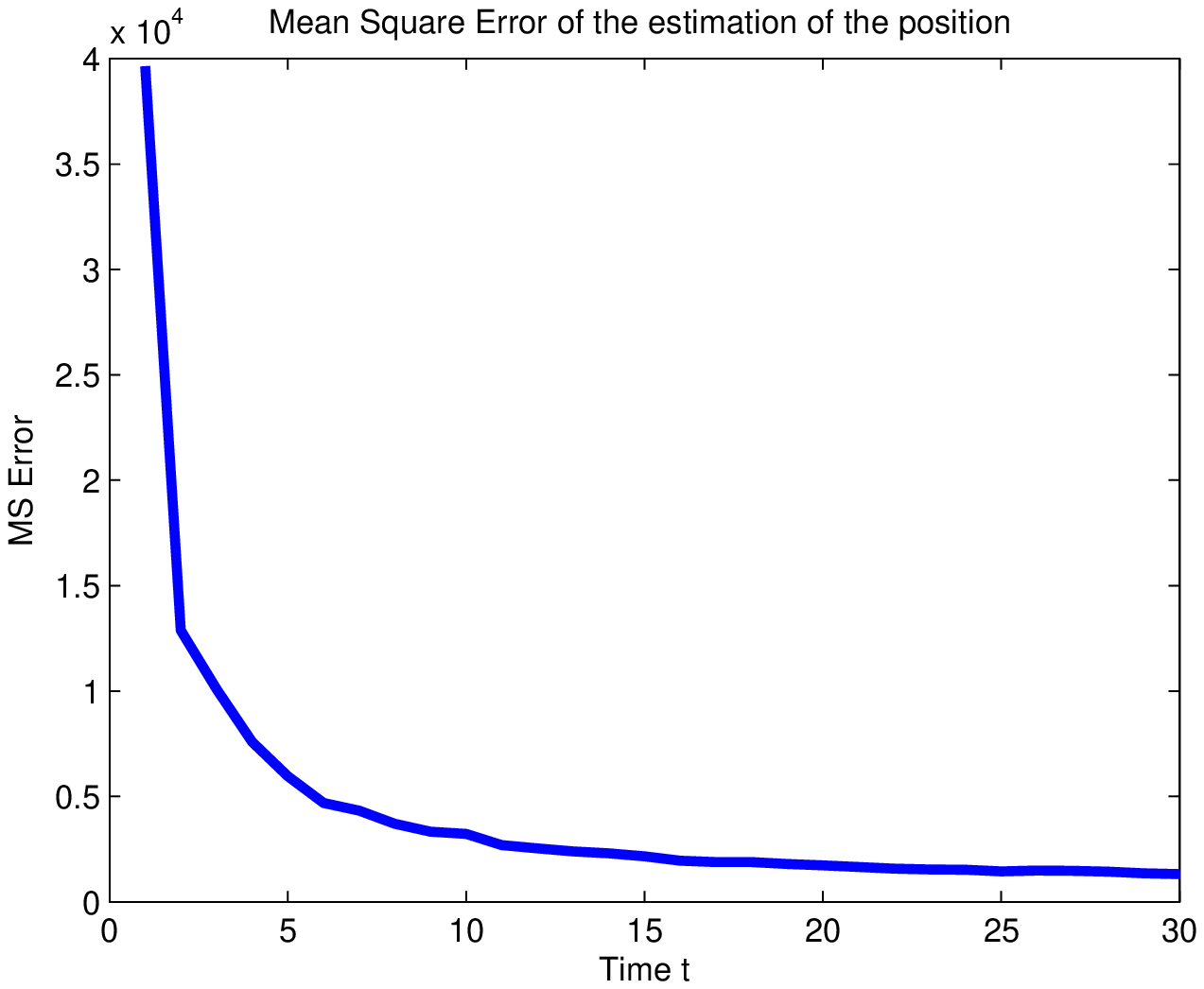}}
{\includegraphics[width=9cm, height=8cm]{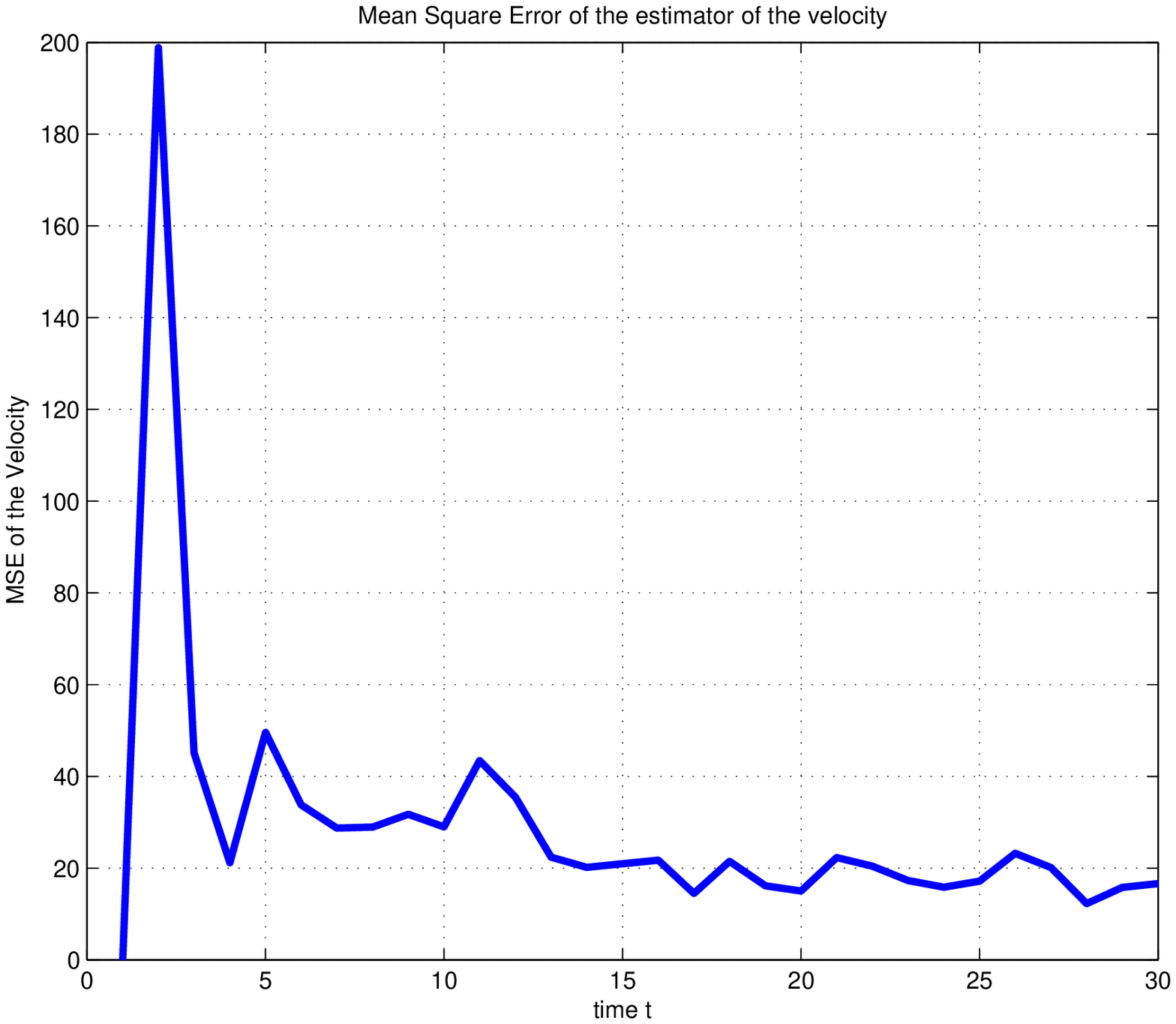}}
\caption{\it MSE of the estimator of the target position. MSE of the estimator of the target velocity.}
\label{mse_state}
\end{figure}
In the same way, the velocity estimation has some acceptable MSE through the tracking process. Despite the clearly strong decrease at the beginning, the curve then stands to an acceptable but non zero value. The effect is more obvious than in the position case, surely because of the velocity modelling we make in the algorithm, which forces the velocity estimation to very bad evolution. A clue could be to perform a most sophisticated modelling of the velocity, but given the binary information, this won't be easy. This is another work in progress for the evolution of our algorithm. 

\section{Conclusion}

\subsection{CVM}

In this paper, we chose to focus on the use of the $\{-,+\}$ at the level of information processing for a sensor network. Though this information is rather poor, it has been shown that it can provide  very interesting results about the  target velocity estimation. The theoretical aspects of our methods have been thoroughly investigated, and it has been shown that the PPR method leads to the right velocity plane  if the number of sensors increase to infinity.  The feasibility of the new concept  ("velocity plane") for estimating the target trajectory parameters  has been put in evidence. The proposed methods seem to be sufficiently general and versatile to explore numerous extensions like: target tracking and dealing with multiple targets within the same binary context.

\subsection{Random walk}

A new method for tracking  both position and velocity of a moving target via  binary data has been developed. Though the instantaneous data are poorly informative, our algorithm takes benefit of the network extent and density via specific spatio-temporal analysis. This is remarkable since the assumptions we made about target motion are not restrictive. Noticeably also, our algorithm is quite fast and reliable. 
Furthermore, it is clear that performance can be greatly improved if we can consider that the acquisition frequency is (far) greater than the maneuver frequency. In particular, we can mix the present method with the one we developed in \cite{aijplc}.\\

However, some important questions remain. The first one concerns the velocity modeling. We focused on this paper on the adaptability of the different correction factors, but we didn't pay much attention to that modeling, which can definitely improve the estimation quality. Moreover, our tracking algorithm is basically deterministic even if the target motion modeling is basically probabilistic. 
Thus, it should be worth to calculate the first correction factor ($\lambda$) via a likelihood, such that ${\bf x}^{corr}$ does not always stays in the mean of the special set. Moreover, that likelihood should be related to all the sources of sensor uncertainty.  In addition, the present algorithm gives a slow response to sudden target maneuver. A remedy should be to incorporate a stochastic modeling of such event in our algorithm.\\

The second correction factor ($\theta$) may also be improved via a stochastic approach. Instead of considering a correction  only related to  the estimated velocity estimated, we could immerse this correction within a stochastic framework involving both ${\hat{\bf v}}_{t}$  and $\theta$. These observations are part of our next work on that very constrained but also quite exciting tracking framework. 
The last important point is  multiple target tracking. Even if our work in this area is quite preliminary, it is our strong belief  that our spatio-temporal separation based algorithm should be the natural way to overcome the association problems.

\end{document}